\documentclass[reprint,nofootinbib,amsmath,amssymb,aps,physrev,journal=prd]{revtex4-2}

\pdfoutput=1
\usepackage{graphicx}
\usepackage{dcolumn}
\usepackage{bm}
\usepackage{hyperref}
\usepackage{lipsum}

\usepackage{subfigure}
\usepackage{calrsfs}

\usepackage{xcolor}

\usepackage{amsthm}

\usepackage{blindtext}
\usepackage{ragged2e}

\usepackage{calrsfs}

\DeclareMathAlphabet{\pazocal}{OMS}{zplm}{m}{n}
\usepackage[scr=boondox]{mathalfa}

\usepackage{xcolor}

\usepackage{amsthm}

\newtheorem*{corollary*}{Corollary}

\begin{document}

\title{Toward a realistic Buchdahl limit in $f(R)$ theories of gravity}

\author{Ra\'ul Carpio Fern\'andez}
    \email{raul.carpio@usal.es}
    \affiliation{ Departamento de F\'{i}sica Fundamental, Universidad de Salamanca, 37008 Salamanca, Spain}
\author{\'{A}lvaro de la Cruz-Dombriz}
    \email{alvaro.dombriz@usal.es}
    \affiliation{ Departamento de F\'{i}sica Fundamental, Universidad de Salamanca, 37008 Salamanca, Spain}
    \affiliation{Cosmology and Gravity Group, Department of Mathematics and Applied Mathematics, University of Cape Town, Rondebosch 7700, Cape Town, South Africa}

\date{\today}

\begin{abstract}
The so-called Buchdahl limit is not yet fully understood in the context of theories of gravity beyond the Einsteinian framework. In this investigation, we generalize this limit for the case of static, spherically symmetric, relativistic compact stars in $f(R)$ theories of gravity within the metric formalism. We present a comprehensive analysis that ensures regularity, thermodynamic stability, fulfillment of all required junction conditions, recovery of the Newtonian potential at long distances and a correct extraction of the asymptotic mass. Our results are exemplified for the $f(R)=R+\alpha R^2$ Starobinsky model and several realistic equations of state describing neutron-star matter. We also compare these results with the case of compact stars immersed \textit{ad hoc} in a Schwarzschild exterior vacuum, although this scenario does not fulfill all necessary $f(R)$ junction conditions. 
To test the validity of viable $f(R)$ models, we show that such stars can indeed host additional energetic content, so their gravitational redshift can be greater than 2, which is prohibited in general relativity. 
\end{abstract}

\maketitle

\section{Introduction}

As widely known in general relativity (GR), when we consider a realistic static and spherically symmetric object embedded in vacuum, an arbitrary relation between its mass and radius is not possible. The reason is that in this case, the Schwarzschild metric is the only possible exterior solution, which exhibits an event horizon at $r_{\rm BH}=2M$, known as the Schwarzschild radius, being $M$ the mass of the object.
Therefore, any object whose radius is smaller than $2M$ is forced to be a black hole (BH). Consequently, any star must have a radius $r_b>2M$. 
%\\
Furthermore, in these spacetimes another restriction appears for the mass-radius relationship, known as the Buchdahl limit. This limit establishes that $M < (4/9)r_b$, which is more restrictive than the Schwarzschild radius. Consequently, a forbidden range in GR for stable stars radii would be $2M < r_b < (9/4)M$.
Apart from being accomplished for all physically viable matter equations of state (EOS), this result can be proved using matching (also dubbed junction) conditions at the boundary of the star, where the spacetime is smoothly matched to the Schwarzschild exterior ({\it cf.} %\cite{padmanabhan2010gravitation}  and 
\cite{
Lake:2017wlx,
bonnor1981junction,
israel1966singular,
darmoisequations} for details on the so-called Darmois-Israel (GR) junction conditions). A direct consequence of the Buchdahl limit is that the gravitational redshift $z$ at the stellar surface, which in GR can be shown to yield
\begin{equation}\label{eq:redshift}
    z = \dfrac{1}{\sqrt{1-\dfrac{2M}{r_b}}} - 1\,,
\end{equation}
is bounded from above ($z\leq 2$).

Recently, some observations have pointed toward the violation of such a limit, either in individual configurations \cite{Karkevandi:2024vov}, by means of gravitational-wave signals from coalescence objects \cite{LIGOScientific:2024elc}, or
when either the monotonicity assumption is relaxed or the energy density is allowed to become negative \cite{Arrechea:2024vxp}.
As such, when considering theories of gravity beyond the Einsteinian paradigm, it is natural to wonder whether the corresponding Buchdahl limit, if existing, can differ from GR predictions. In other words, whether the mass-radius relationship of a compact star can fall within the GR forbidden region, which as a byproduct would imply that the gravitational redshift of (some) stars would be greater than 2. 
The resolution of this problem is of great importance, since should the fact that the gravitational redshift of a compact star can be greater than 2 be experimentally demonstrated, it would serve as an experimental test to validate modified theories in the strong-gravity regimes. In fact, the gravitational redshift can be measured through terrestrial, solar system, and astronomical observations. For example, given the composition of the stars, it can be calculated by analyzing the redshift of the spectral lines \cite{2010JHA....41...41H}. Another possibility consists in measuring this quantity in the Sun---and thus applicable to other stars---using optical methods, as detailed in \cite{1962PhDT........57B}.

In order to shed some light on this open problem, we have decided to work with the paradigmatic scalar-tensor $f(R)$ gravity theories in the metric formalism ({\it cf.} \cite{Sotiriou:2008rp,DeFelice:2010aj} for the foundations of such theories). The static and spherically symmetric vacuum solutions for such theories have been widely studied \cite{Clifton:2006ug,delaCruz-Dombriz:2009pzc,hurgobin2019stability,Nzioki:2009av,Casado-Turrion:2023rni,Olmo:2019flu,Calza:2018ohl,Campbell:2024ouk}, as well as the phenomenology of compact objects \cite{Astashenok:2017dpo,
AparicioResco:2016xcm} and the gravitational collapse therein \cite{Goswami:2014lxa,Casado-Turrion:2022xkl,Cembranos:2012fd,Vinckers:2023oxa,Astashenok:2018bol}. Due to the fact that in the context of $f(R)$ theories both the Jebsen-Birkhoff theorem breaks down and the junction conditions gluing two spacetime patches are more restrictive than their GR counterparts ({\it cf.} the elegant derivation for $f(R)$ theories in \cite{Senovilla:2013vra,Deruelle:2007pt}), the derivation of a Buchdahl limit remains a challenging issue to be addressed in the following.
 
The article is thus organized as follows: In Sec. \ref{Sec:II} we shall first introduce the $f(R)$ field equations in the metric formalism, which are usually integrated to solve both the interior of the star, endowed with realistic EOS, and the matter vacuum exterior. 
%

%Therein, \ref{Sec:IIA} shall be devoted to present the complete system of equations to solve and the necessary initial and asymptotic conditions to be satisfied.
%
Therein, Sec. \ref{Sec:IIA} deals with the stability conditions to be satisfied by the solutions, followed in Sec. \ref{Sec:IIB} by the discussion of the usual metric $f(R)$ junction conditions to be imposed at the edge between inner and outer spacetimes. Subsequently, Sec. \ref{Sec:IIC} presents the methodology to find the numerical solutions 
and provides one explicit realization of our code for the $f(R)$ quadratic Starobinsky model \cite{Starobinsky:1980te}. At that stage, we shall briefly mention the EOS to be adopted throughout this investigation for the matter content (saliently the neutron fluid) in the star.
To conclude that section, in Sec. \ref{Sec:IID} we tackle the required definition of the gravitational mass as perceived by an asymptotic observer.
Once the required foundations are exposed, Sec. \ref{Sec:IIIA} shall be devoted to find the generalization of the Buchdahl limit in generic $f(R)$ theories
%. In order to do so we have advanced through inequalities which are satisfied for all generic $f(R)$ theories 
satisfying the viability conditions presented in the previous section. As an application, % illustrating the power of our results, 
in Sec. \ref{Sec:IIIB} we shall apply our results again to the quadratic Starobinsky $f(R)$ model for a wide class of realistic EOS.
For comparison purposes in Sec. \ref{Sec:IIIC} we obtain a kind of Buchdahl limit again in generic $f(R)$ theories although imposing the outer spacetime to be purely Schwarzschild which, despite its elegance, cannot be matched with $f(R)$ interior solutions in general.
Immediately after, in Sec. \ref{Sec:IVA}, we shall present the most relevant implications that the obtained results may have 
%in the correct interpretation of results 
depending on whether the assumed outer spacetime is 
a Schwarzschild patch or
the one gluing smoothly to realistic interior solutions. Herein we shall also present the upper bounds for the mass increment when comparing the maximum achievable mass in $f(R)$ with the GR counterpart. Then, in Sec. \ref{Sec:IVB} we shall present specific results for the mass upper limit, the mass increment and the gravitational redshifts for the family of $f(R)$ models under study.
Finally, a discussion and prospects of the results are provided in Sec. \ref{Sec:V}. 
Throughout this investigation, we used geometrized units $c=G=1$. In addition, the metric signature is +2.\\

The interested reader is referred to the technical calculations presented in the Appendixes. 
%Their specific contents are described at the beginning of this section.  
Appendix \ref{App:apend_A} includes the derivation of the pertinent equations of motion and a detailed explanation of the chosen initial conditions. Then, Appendix \ref{App:apend_B} displays the method to find asymptotically flat solutions. Next, in Appendix \ref{App:apend_C} we present the formalism we use to obtain the expression for the Buchdahl limit \eqref{eq:Buch_MfR}. Finally, Appendix \ref{App:apend_D} provides some crosschecks to guarantee the upper-bound ansatz \eqref{eq:Buch_funcional} made for quadratic Starobinsky $f(R)$ models, which relates the asymptotic mass and the mass function when evaluated at the radius of the star.

\section{Static and spherically symmetric solutions in $f(R)$ theories}
\label{Sec:II}
The total action in $f(R)$ theories of gravity is given by
\begin{equation}
    S = \dfrac{1}{2\kappa} \int {\rm d}^4x \sqrt{-g}\left[f(R) + \mathcal{L}_M(g_{\mu\nu},\phi) \right]\,,
\end{equation}
where $\kappa\equiv 8\pi$, $g$ is the metric determinant and $\mathcal{L}_M$ corresponds to the Lagrangian associated with matter fields $\phi$. By varying the action with respect to the metric, 
the field equations in the metric formalism are obtained, yielding
\begin{equation}\label{eq:field_ecs}
    R_{\mu\nu}f_R - \dfrac{1}{2}g_{\mu\nu}f(R)+\left( g_{\mu\nu}\Box - \nabla_\mu\nabla_\nu \right)f_R  = \kappa T^M_{\mu\nu}\,,
\end{equation}
where $f_{R}={\rm d}f(R)/{\rm d}R$, $\square\equiv\nabla^\mu\nabla_\mu$ and $T^M_{\mu\nu}$ is the energy-momentum tensor associated with matter ($M$), which is defined by
\begin{equation}
    T^M_{\mu\nu} = -\dfrac{2}{\sqrt{-g}}\dfrac{\delta \mathcal{L}_M}{\delta g^{\mu\nu}}\,.
\end{equation}
As widely known, the field equations \eqref{eq:field_ecs} can be written in the {\it \`{a} la} Einstein as follows:
\begin{equation}\label{eq:field_eqs_effective}
G_{\mu\nu} 
\equiv R_{\mu\nu}-\dfrac{1}{2}g_{\mu\nu}R 
= \kappa \left( \tilde{T}^M_{\mu\nu} + T_{\mu\nu}^{R} \right) \equiv \kappa\,T_{\mu\nu}\,,
\end{equation}
where $T_{\mu\nu}$ denotes the total energy-momentum tensor, composed of an ``effective matter'' contribution,
\begin{equation}
\tilde{T}^M_{\mu\nu} \equiv \dfrac{T^M_{\mu\nu}}{f_R}\,,
\end{equation}
and a contribution associated with a ``curvature'' fluid ($R$), 
\begin{equation}\label{eq:energy_mom_curvature}
T_{\mu\nu}^{R} \equiv \dfrac{1}{\kappa f_R} \left[ \dfrac{1}{2} g_{\mu\nu}\left( f(R) - Rf_R \right)+ \left( \nabla_\mu\nabla_\nu - g_{\mu\nu}\Box \right)f_R \right]\,,
\end{equation}
the latter vanishing for $f(R)=R$, i.e., in the GR scenario. %This procedure may seem questionable and unnatural, as it is not the theory of Einstein and we are forcing it to have an interpretation in terms of Einstein's equations. However, 
As we shall see later in this section, the unnatural decomposition \eqref{eq:field_eqs_effective} will prove to be useful.

Since we are interested in spherically symmetric and static configurations, the pertinent metric can be expressed as
\begin{equation}\label{eq:metric}
    {\rm d}s^2 = -B(r)\,{\rm d}t^2 + A(r)\,{\rm d}r^2 + r^2{\rm d}\Omega^2 \,,
\end{equation}
where $A$ and $B$ are at least $C^2$ functions.
Also, we shall assume that the matter content in the interior of stars under consideration can be described as a perfect fluid, so its energy-momentum tensor becomes
\begin{equation}\label{eq:perfectfluid}
    T^{M}_{\mu\nu} = (\rho + p)u_\mu u_\nu + p\,g_{\mu\nu}\, .
\end{equation}
Also, for the choice of the metric \eqref{eq:metric}, the total energy-momentum tensor can be expressed as
\begin{equation}\label{eq:Buch_8}
T^{\mu}_{\;\;\nu}=\operatorname{diag}\left[-\rho_{total}(r), p_r(r), p_\theta(r), p_\theta(r)\right]\,,
\end{equation}
where we have defined
\begin{align}
\nonumber
& \rho_{total}(r)=\dfrac{\rho(r)}{f_{R}}+\rho^R(r) \,,\\ \label{eq:rho_p_(r)}
& p_r(r)=\dfrac{p(r)}{f_{R}}+p_r^R(r) \,,\\ \nonumber
& p_\theta(r)=\dfrac{p(r)}{f_{R}}+p_\theta^R(r) \,,
\end{align}
separating the corresponding contributions from matter and curvature.\footnote{$\rho^R(r)$, $p_r^R(r)$ and $p_\theta^R(r)$ in \eqref{eq:rho_p_(r)} can be computed just by evaluating the pertinent components of \eqref{eq:energy_mom_curvature}.} This concept will be useful in Sec. \ref{Sec:III} below.
Now, substituting expressions \eqref{eq:energy_mom_curvature}--\eqref{eq:perfectfluid} in the field equations \eqref{eq:field_eqs_effective}, and using the conservation of the energy-momentum tensors $T_{\mu\nu}$ and $T^{M}_{\mu\nu}$, i.e., $\nabla^\mu T_{\mu\nu}=0=\nabla^\mu T^{M}_{\mu\nu}$, we obtain the following system of coupled ordinary differential equations (full derivation for the interested reader is provided in Appendix \ref{App:apend_A}):
\begin{gather}\nonumber
    A' = \dfrac{2rA}{3f_R}\left[ \kappa A (\rho + 3p) + Af(R) - f_R\left( \dfrac{AR}{2} + \dfrac{3B'}{2rB} \right)\right. \\ \label{eq:1edo}
    \left. - \left( \dfrac{3}{r} + \dfrac{3B'}{2B} \right)f_{2R}R' \right] \,,
\end{gather}
\begin{gather}\nonumber
    B'' = \dfrac{B'}{2}\left( \dfrac{A'}{A} + \dfrac{B'}{B}\right) + \dfrac{2A'B}{rA} + \dfrac{2B}{f_R}\left[ -\kappa A p \right.\\ \label{eq:2edo}
    \left. + \left( \dfrac{B'}{2B}  + \dfrac{2}{r}\right)f_{2R}R' - 
\dfrac{Af(R)}{2} \right] \,,
\end{gather}
\begin{gather}\nonumber
    R'' = R'\left( \dfrac{A'}{2A} - \dfrac{B'}{2B} - \dfrac{2}{r}\right) - \dfrac{f_{3R}R'^2}{f_{2R}} \\ \label{eq:3edo}
    - \dfrac{A}{3f_{2R}}\left[ \kappa(\rho-3p) + f_R R - 2f(R)\right]  \,,
\end{gather}
\begin{gather}\label{eq:4edo}
    p' = -  \dfrac{\rho + p}{2}\dfrac{B'}{B} \,,
\end{gather}
where the symbol $'$ denotes a derivative with respect to the radial coordinate $r$ and $f_{(i)R}={\rm d}^{(i)}f(R)/{\rm d}R^{i}$ with $i=1,2,3$.

An alternative parametrization of the metric \eqref{eq:metric} involves introducing the functions 
\begin{equation}\label{eq:Buch_metric1}
c(r)\equiv \sqrt{B(r)}\,, \quad m(r)\equiv \frac{r}{2}(1-1/A(r))\,.
\end{equation}
As we can see in \eqref{eq:rho_p_(r)}, when considering a perfect fluid, there is no pressure anisotropy associated with matter, that is, the contribution from matter in both $p_r(r)$ and $p_\theta(r)$ is the same. However, this is not the case should $f(R)$ curvature terms be present. More specifically, using the definition of $T_{\mu\nu}^R$ according to \eqref{eq:energy_mom_curvature} and the metric of interest once parametrized as per \eqref{eq:Buch_metric1}, the pressure anisotropy associated with curvature yields
\begin{gather}\nonumber
p_\theta^R-p_r^R = \dfrac{1}{\kappa f_{R}}\left\{\left(\dfrac{m'}{r}-\dfrac{m}{r^2}\right) R' f_{2R} \right.\\\label{eq:anisotropy}
\left. -\left(1-\dfrac{2 m}{r}\right)\left[\left(R' f_{2R}\right)'-f_{2R} \dfrac{R'}{r}\right]\right\}\,.
\end{gather}
On the other hand, from the field equation $G^0_{\;\;0} = \kappa\, T^0_{\;\;0}$ in \eqref{eq:field_eqs_effective}, we conclude that
\begin{equation}
\label{eq:Buch_mu}
\rho_{total} = \dfrac{2}{\kappa}\dfrac{m'}{r^2} \Longrightarrow m(r) = \dfrac{\kappa}{2}\int_0^r \rho_{total}(x)\,x^2{\rm d}x\,.
\end{equation}
Since in the context of $f(R)$ the exterior metric matching suitable interiors usually does not coincide with the %necessarily 
Schwarzschild one, the value of $m(r_b)$ does not need to coincide with the Schwarzschild asymptotic mass, as will be shown below. 
Consequently, according to \eqref{eq:Buch_mu} we can just interpret the function $m(r)$ as the effective mass that is generated by both matter and curvature fluids enclosed within a sphere of radius $r$. 
%\textcolor{blue}{{[\bf Alvaro - quitar]} At the star's surface, $m(r_b)$ represents the total effective mass of the star. whose exterior region would correspond - even violating the aforementioned junction conditions - to the Schwarzschild mass.} 

The remaining equations of motion in \eqref{eq:field_eqs_effective} relate the effective mass $m(r)$ to the pressure terms. For instance $G^1_{\;\;1} = \kappa\, T^1_{\;\;1}$ yields
\begin{equation}\label{eq:Buch_pr}
p_r=\dfrac{2 c'}{\kappa\, r c}\left(1-\dfrac{2 m}{r}\right)-\dfrac{2 m}{\kappa\,r^3}\,.
\end{equation}
Furthermore, using the second Bianchi identity $\nabla_{\alpha} G^{\alpha}_{\;\;r}=0=\nabla_{\alpha} T^{\alpha}_{\;\;r}$, we arrive at the following expression:
\begin{equation}\label{eq:Buch_cpr}
\left(c p_r\right)'+c' \rho_{total}=\dfrac{2 c}{r}\left(p_\theta-p_r\right)\,.
\end{equation}
Now, substituting \eqref{eq:Buch_mu} and \eqref{eq:Buch_pr} into the left-hand side of \eqref{eq:Buch_cpr} and simplifying, we obtain
\begin{equation}\label{eq:Buch_13}
\sqrt{1-\dfrac{2 m}{r}} \dfrac{{\rm d}}{{\rm d} r}\left(\dfrac{c'}{r} \sqrt{1-\dfrac{2 m}{r}} \right)=c\left[\left(\dfrac{m}{r^3}\right)'+\dfrac{\kappa(p_\theta-p_r)}{r}\right]\,.
\end{equation}
Equation \eqref{eq:Buch_13} is often used in the literature to study modifications of the Buchdahl limit in the case of stars with pressure anisotropies within the GR context. However, herein the anisotropies as in \eqref{eq:anisotropy} emerge due to the introduction of $f(R)$ curvature terms in the total energy-momentum tensor. 
Thus, since there are no anisotropies associated with matter, a simple inspection of \eqref{eq:rho_p_(r)} leads us to conclude that $p_\theta-p_r=p_\theta^R-p_r^R$. Therefore, the combination of Eqs. \eqref{eq:anisotropy} and \eqref{eq:Buch_13} provides
\begin{align}\nonumber
    & f_{R} \dfrac{{\rm d}}{{\rm d} r}\left(\dfrac{c'}{r} \sqrt{1-\dfrac{2 m}{r}} \right)
    +c \dfrac{{\rm d}}{{\rm d} r}\left(\dfrac{f_R'}{r} \sqrt{1-\dfrac{2 m}{r}}\right)\\ \label{eq:cond_17}
    & =\dfrac{c f_{R}}{\sqrt{1-\dfrac{2 m}{r}}} \dfrac{{\rm d}}{{\rm d} r}\left(\dfrac{m}{r^3}\right)\,.
\end{align}
%As mentioned earlier, the term $m/r^3$ on the right-hand side may be understood as the effective average density of the star at radius $r$. 
On the left-hand side of the above expression, we can notice that the functions $c(r)$ and $f_R$ play a symmetrical role.
Additionally, when $f_R\equiv 1$, i.e., in the GR context, the expression \eqref{eq:cond_17} once simplified can be directly used to obtain the usual Einsteinian Buchdahl limit.

%As mentioned above, in order to solve the system of equations within compact objects, we also need an EOS relating matter density and matter pressure. This relationship crucially depends upon the type of object under study, herein neutron stars throughout this investigation.
%
%For these objects, it is common the use of numerical EOS, derived from particle-physics models and observational results. In the following we shall use three paradigmatic EOS presented in Figure \ref{fig:EOS_neutronstar}, labeled as ``Soft'', ``Middle'' and ``Stiff''. 

\begin{figure}
    \centering
\includegraphics[width=0.4\textwidth]{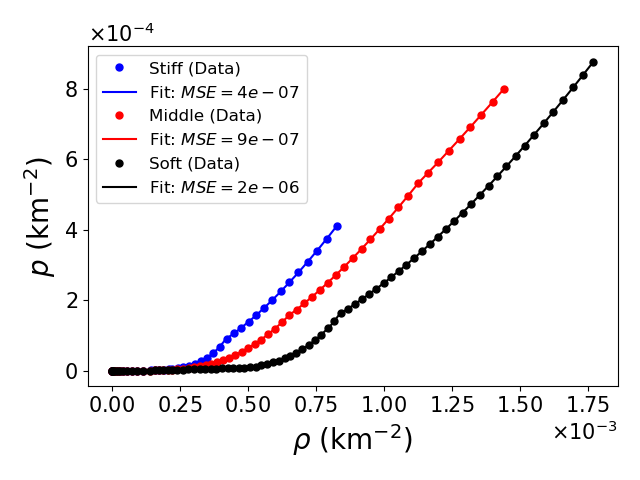}
    \caption{``Stiff'', ``middle'' and ``soft'' EOS based on potential models' data as in \cite{Hebeler:2013nza} that describe neutron matter, and their respective fits. Soft possesses an EOS in which pressure increases most slowly with density, and stiff is the one in which this growth is the most rapid. The error in the spline fits was estimated with the usual mean-squared errors (MSEs).}
\label{fig:EoS_neutronstar}
\end{figure}

\subsection{Viability and regularity conditions}
\label{Sec:IIA}

To construct a stable (in terms of thermodynamics) and continuous stellar model that can be matched with an empty exterior, it is required to impose certain constraints on both the different thermodynamic quantities of the star and the interior metric components. Additionally, to avoid the presence of ghosts or tachyons and to ensure the attractive nature of gravity together with the avoidance of the well-known Dolgov-Kawasaki instability, the usual conditions on the function $f(R)$,
\begin{equation}\label{eq:cond_18}
f_R > 0 \quad \text{and} \quad f_{2R} \leq 0\,,
\end{equation}
must be obeyed.\footnote{
When conditions \eqref{eq:cond_18} apply to $f(R)=R+\alpha R^2$ models, well-behaved solutions require $\alpha<0$. For different choices of the metric signature and the Riemann tensor definition, the condition on the sign of $f_{2R}$ may be the opposite one.}
%

%\vspace{0.1cm}
% \subsection{Regularity conditions and thermodynamic stability}
% This is III.C. original, now moved here
%
%{\it Regularity conditions}: 
Then, given that the components of the metric and thermodynamic functions must be at least of class $C^2$ within the star, the derivatives with respect to the radial coordinate of these functions at the center of the star must vanish, i.e.,
\begin{equation}\label{eq:cond_23}
    p'(0)=\rho'(0)=R'(0)=0\,.
\end{equation}
Moreover, at the center of the star we consider $B'(0)=0$, which can be interpreted either as a regularity condition or an initial condition. Moreover, in the interior of a star the condition 
\begin{equation}\label{eq:regul_B}
    B'(r) \geq 0 \quad \text{with} \quad \ 0 \leq r \leq r_b
\end{equation}
needs to be satisfied. This can be interpreted as $B(r)$ being related to the gravitational potential, so for the star to be thermodynamically stable, this function must not have any local extremum at any point other than the center. Thus, $B(r)$ is a monotonically increasing and positive function in the interior of the star.\footnote{The regularity condition \eqref{eq:regul_B} in the parametrization \eqref{eq:Buch_metric1} yields
\begin{equation}\label{eq:cond_25}
    c'(r) \geq 0, \quad \ 0 \leq r \leq r_b \quad \text{with} \quad c'(0)=0\,,
\end{equation}
so that $c(r_b)\geq c(r)>0$, $\forall r\in [0,r_b]$.}

\begin{figure}
    \centering
    \subfigure{
        \includegraphics[width=0.23\textwidth]{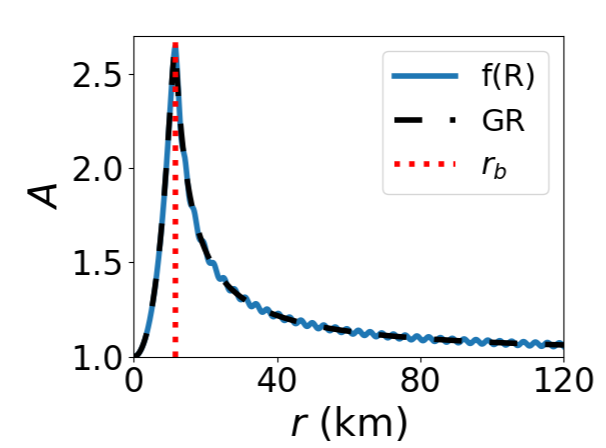}}
    \subfigure{
        \includegraphics[width=0.23\textwidth]{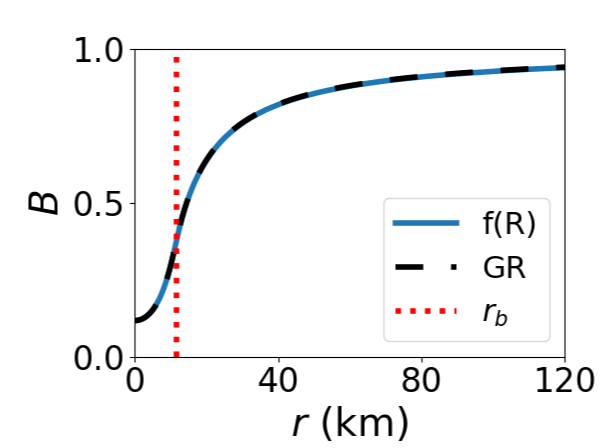}}
    \subfigure{
        \includegraphics[width=0.23\textwidth]{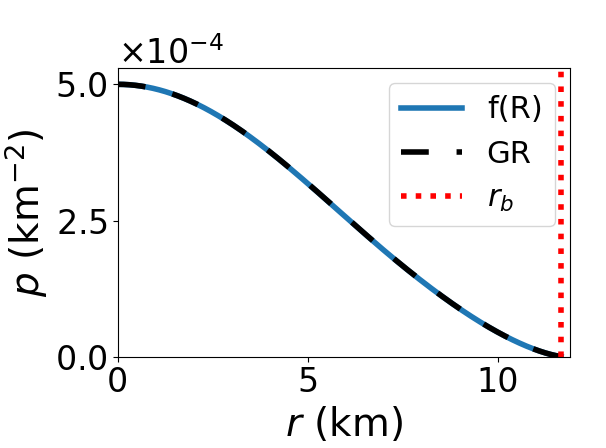}} 
    \subfigure{
        \includegraphics[width=0.23\textwidth]{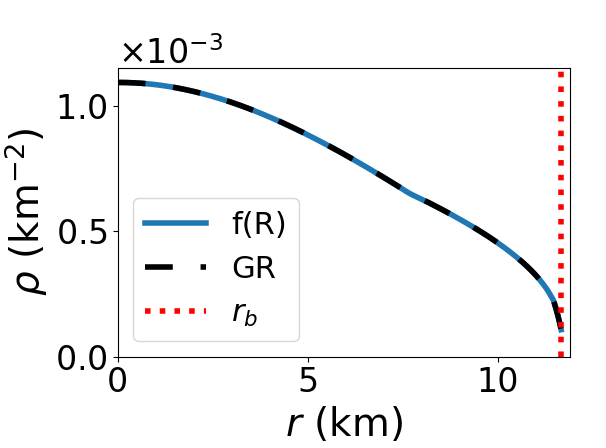}} 
    \subfigure{
        \includegraphics[width=0.24\textwidth]{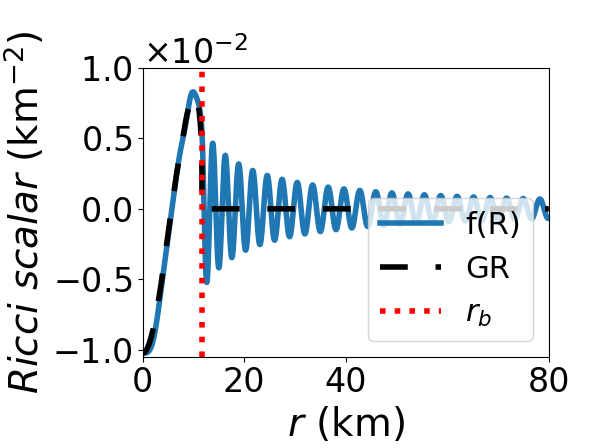}} 
    \caption{Numerical solutions of the field equations for both GR and $f(R)=R+\alpha R^2$ with $\alpha=-0.05\, {\rm km}^2$. Clockwise from top left, $A(r)$, $B(r)$, $\rho(r)$, $R(r)$ and $p(r)$. For a central pressure $p_c=5\cdot 10^{-4}\,{\rm km}^{-2}$ and the middle EOS, a radius $r_b=11.670\,{\rm km}$ was obtained for the $f(R)$ model, whereas $r^{\rm GR}_b = 11.669\,{\rm km}$ for such EOS and same initial conditions. In $f(R)$ theories, the functions $A(r)$ and $B(r)$ exhibit the same trend as their GR counterparts, although the former show damped oscillations in the exterior of the star. Such oscillations of $B(r)$ are considerably smaller and cannot be appreciated in this example. The curvature scalar $R(r)$ also exhibits damped oscillations in the exterior, in contrast with GR, where $R$ is identically zero once in (Schwarzschild) vacuum.}
    \label{fig:Buch_comparation}
\end{figure}

Moreover, as discussed for instance in \cite{ferrari2020general}, for a star to be physically realistic, we assume that the energy density and pressure must be positive throughout the star, i.e.,
\begin{equation}\label{eq:cond_p_rho}
    \rho \geq 0, \quad p \geq 0\,.
\end{equation}
As a natural generalization from the GR counterpart, the total density needs to be a monotonically decreasing function of $r$, ergo
\begin{equation}\label{eq:cond_26}
    \dfrac{{\rm d}}{{\rm d}r}\left( \dfrac{m}{r^3} \right) \leq 0\,, 
\end{equation}
the latter being necessary to preserve the stability of the star. In fact, as we shall see in Fig. \ref{fig:Buch_comparation}---as well as all the other test simulations for different EOS which we have run---the conditions \eqref{eq:regul_B} and \eqref{eq:cond_p_rho} are automatically satisfied as a consequence of the numerical resolution of the field equations, so no prior imposition on the shapes of $B(r)$, $\rho(r)$ or $\rho_{total}(r)$ is needed. This is a by-product result of our numerically exact analysis which remained unclear in previous literature.

\subsection{Junction conditions}
\label{Sec:IIB}
Following the notation in \cite{poisson_2004}, in order to study a compact star surrounded by matter vacuum, we consider two spacetimes $V^+$ and $V^-$ of class $C^3$ with their respective metrics $g^+$ and $g^-$ of class $C^2$, and boundaries $\Sigma^+$ and $\Sigma^-$ (timelike type).\footnote{
$V^+$ and $V^-$ are assumed to be $C^3$-class (smooth up to the third derivative) to ensure sufficient smoothness for solving the field equations. 
%The condition $C^2$ (differentiable up to second order) for $g^+$ and $g^-$ ensures the continuity of the metric and its first derivatives. 
%
$\Sigma^+$ and $\Sigma^-$ are both considered time-like to ensure the proper causal structure.
} We will consider a single hypersurface $\Sigma$ that separates the two regions $V^+$ and $V^-$. 

As carefully explained in \cite{Senovilla:2013vra}, in the context of metric $f(R)$ gravity, the junction conditions for a compact star---excluding the existence of thin shells, double layers, etc.---impose the continuity of both $g_{\alpha\beta}$, $K_{\alpha\beta}$ (that is, the extrinsic curvature), $R$ and $\nabla_\alpha R$. Hence,\footnote{To study the conditions that must be satisfied at $\Sigma$, the following notation is introduced:
\begin{equation}
    [A] \equiv A(V^+)|_\Sigma - A(V^-)|_\Sigma\,,
\end{equation}
where $A$ can be any tensorial quantity defined in both regions.}
\begin{equation}
    [g_{\alpha\beta}] = 0\,,\;
    [K_{\alpha\beta}] = 0\,,\;
    [R] = 0\,,\;
    [\nabla_\alpha R] = 0\,.
    \label{junction_conditions_fR}
\end{equation}
Due to the last two equalities, it also must be fulfilled that 
\begin{equation}\label{eq:junction_noshell_5}
    n^\alpha[T_{\alpha\beta}]=0\,,
\end{equation}
where $n^\alpha$ represents a vector orthogonal to the hypersurface $\Sigma$. In the case of a spherical static star endowed with standard matter modeled as a perfect fluid, the last condition yields
\begin{equation}
\label{matter_rho_P_at_rb}
p(r_b)=0\,,\;\rho(r_b)=0\,,
\end{equation}
which serves to compute the radius star. Note that the density and pressure contributions issued from the {\it curvature} fluid do not vanish at the radius star undoubtedly.

Also, as mentioned in \cite{Senovilla:2013vra}, in the specific case of a Schwarzschild exterior it is also required that
\begin{equation}\label{eq:junction_cond_Schw_prho}
%\dfrac{{\rm d}p}{{\rm d}r}(r_b)=0\,,\;
%\dfrac{{\rm d}\rho}{{\rm d}r}(r_b)=0.
p'(r_b)=0\,,\;
\rho'(r_b)=0.
\end{equation}
Due to the above conditions, even though in $f(R)$ there may exist static and spherically symmetric interior solutions that can be matched with a Schwarzschild exterior, the perfect fluid solutions that one finds in GR---forced to satisfy \eqref{eq:junction_cond_Schw_prho}---will generally not be among them.

\subsection{Numerical resolution of the equations}
\label{Sec:IIC}
In order to find the solutions for $A(r)$, $B(r)$, $R(r)$, $p(r)$ and $\rho(r)$ resorting to Eqs. \eqref{eq:1edo}--\eqref{eq:4edo} above, an EOS $\rho = \rho(p)$ as well as six initial conditions---usually provided at the center of the star, i.e., $\{A(0), B(0), B'(0), R(0), R'(0), p(0)\}$---are required. The three EOS we study in this investigation are summarized in Fig. \ref{fig:EoS_neutronstar}.
Hence, by following the process presented in \cite{AparicioResco:2016xcm}, we consider such conditions to be $A(0)=1$, $B'(0)=0$,\footnote{
These conditions for $A(0)$ and $B'(0)$ translate to $m(0)=0$ and $c'(0)=0$, whereas the value of $c(0)$ is obtained resorting to a shooting method, just as $B(0)$ as explained in Appendix 
\ref{App:apend_B}.
} $R'(0)=0$ and $p(0)=p_c$ (central pressure), while for $R(0)$ we take the corresponding value from GR, i.e., $R(0)=-\kappa\,T^M(0)$, where $T^M=g^{\mu\nu}T^M_{\mu\nu}$ is the trace of the matter contribution of the energy-momentum tensor \eqref{eq:perfectfluid}. 
Regarding the value of $B(0)$, this is derived by using the shooting method as explained in Appendix \ref{App:apend_B}.\footnote{Throughout this investigation, the system of equations \eqref{eq:1edo}--\eqref{eq:4edo} has been solved resorting to a fourth-order Runge-Kutta algorithm.}
Once we specify the initial conditions, we integrate outward in the radial coordinate until the matter pressure vanishes%(or it is sufficiently small)
. At this point, the boundary of the star has been found. From there we set $p=0$, which effectively reduces the system by one differential equation since
\eqref{eq:4edo}
% in the star's exterior region, and so the EOS 
is no longer necessary. At the boundary, we impose the usual $f(R)$ junction conditions as described in Sec. \ref{Sec:IIB} and continue the integration process in vacuum extending outward.

Furthermore, since at very large distances we aim for the metric to behave as that of Schwarzschild, we impose that the found solutions for $A(r)$, $B(r)$ and $R(r)$ must satisfy
\begin{equation}
\label{eq:conditions_schw}    \lim_{r\to\infty}A(r)B(r) = 1 \quad \text{and} \quad \lim_{r\to\infty}R(r) = 0 \,.
\end{equation}
With these asymptotic conditions, we have all the necessary conditions to solve the system of Eqs. \eqref{eq:1edo}--\eqref{eq:4edo} in both the interior and the exterior (there $\rho=p=0$) of the star, while guaranteeing asymptotic flatness.

For a given realization, the numerical solutions are depicted in Fig. \ref{fig:Buch_comparation}.
In this specific case, we have determined that the star's radius is $r_b=11.820\,{\rm km}$ and
visualized that the $A(r)$, $B(r)$ and $R(r)$ functions in the exterior exhibit---
unlike the Schwarzschild solution---an oscillatory behavior. Furthermore, $R(r)$ and $R'(r)$ no longer vanish at the boundary, confirming that the interior solution cannot be smoothly joined with a Schwarzschild exterior should conditions \eqref{junction_conditions_fR} be obeyed. 
We can also verify that, as $r\to\infty$, the functions $A(r)$, $B(r)$ and $R(r)$ satisfy the conditions \eqref{eq:conditions_schw}. On the other hand, both matter pressure and matter density in the interior decrease as the distance to the center of the star increases, while $p(r_b)=\rho(r_b)=0$ at the star radius.

\subsection{Gravitational mass perceived by a far observer}
\label{Sec:IID}
\begin{figure}
    \centering
    \includegraphics[width=0.35\textwidth]{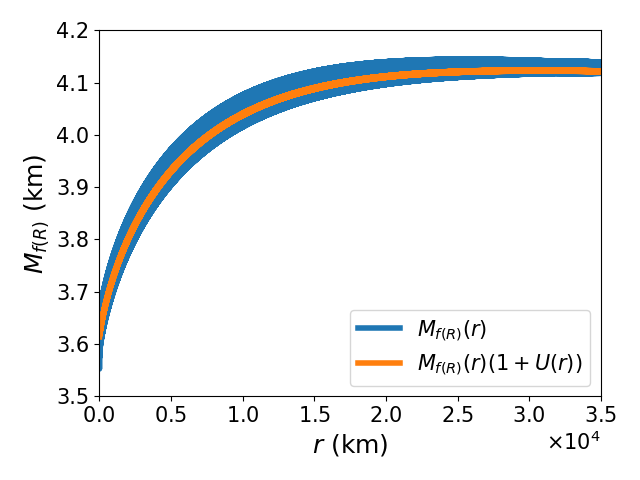}
    \caption{$M_{f(R)}(r)$ and $M_{f(R)}(r)(1+U(r))$ for the $f(R)=R+\alpha R^2$ model with $\alpha=-0.05\, {\rm km}^{2}$. As in Fig. \ref{fig:Buch_comparation}, we have considered $p_c=5\cdot 10^{-4}\,{\rm km}^{-2}$ and the middle EOS. Oscillations in $M_{f(R)}(r)$, whose the characteristic amplitude is smaller than the radial distance scale, appear as a blue band. $M_{f(R)}(r)$ inherited such oscillations from its definition \eqref{eq:metric_functions} in terms of $A(r)$ and $B(r)$.    
    Both functions tend to a constant value $M_{f(R)}^{\infty} = 4.1\, {\rm km}$ at infinity.}
    \label{fig:MfR}
\end{figure}

The definition of mass in $f(R)$ theories is more convoluted than the usual GR counterpart%\footnote{We refer the interested reader to the thorough discussion in \cite{AparicioResco:2016xcm}.}
. 
%
%Indeed, in Newtonian physics, the mass of an object with spherical symmetry is defined as the volume integral of the density of that object. There is of course the mass (energy content) contained within the radius of the star. Complementarily, should the Newtonian potential outside the star is recovered at sufficiently long distances, then the constant on the $-1/r$ term could be assimilated to the asymptotic mass. For a Schwazchild exterior, both notions obviously coincide.
%
%
Indeed, in $f(R)$ theories we can have asymptotically flat spherically symmetric static vacuum solutions that are not exactly Schwarzschild, thus the gravitational potentials being different strictly speaking from the latter. Consequently, following the reasoning in \cite{AparicioResco:2016xcm}, it is useful to introduce the parameterizations
\begin{equation}\label{eq:metric_functions}
    A(r) = \dfrac{1+U(r)}{B(r)}\, , \,
    B(r) = 1 - \dfrac{2M_{f(R)}(r)(1+U(r))}{r}\, ,
\end{equation}
for the metric \eqref{eq:metric}, where $M_{f(R)}(r)$ and $U(r)$ are arbitrary functions that depend solely on the radial coordinate. In this context $U(r)\neq0$ in general and the numerator $M_{f(R)}(r)(1+U(r))$, even in the exterior of the star, is not a constant parameter as in a pure Schwarzschild exterior, but rather a function depending on the radial coordinate. Since we aim to recover the Schwarzschild-like form at infinity, the functions in \eqref{eq:metric_functions} must satisfy, together with the conditions \eqref{eq:conditions_schw},
\begin{equation}
\label{eq:exterior_cond}
\lim_{r\to\infty} U(r) = 0 \quad \text{and}
    \quad \lim_{r\to\infty} \dfrac{M_{f(R)}(r)\left(1+U(r)\right)}{r} = 0\,.
\end{equation}
%
%As discussed in \cite{AparicioResco:2016xcm}, 
When choosing the mass definition in $f(R)$ theories, we must consider that the corrections introduced by these models should be small and limited to the vicinity of compact objects. Thus, when observing a star from a long distance, the usual Newtonian potential ought to be recovered, which allows for the natural interpretation of the total mass of a star. %The effective potential in this case can be shown to be $-M_{f(R)}(r)/r$.
As a result, the mass an observer at infinity would measure is
\begin{equation} \label{eq:M_inf}
    M_{f(R)}^\infty = \lim_{r\to\infty} M_{f(R)}(r)\,.
\end{equation}
By following the numerical procedure developed in Sec. V of \cite{AparicioResco:2016xcm}, we are able to obtain the function $M_{f(R)}(r)$ following the procedure described in Appendix \ref{App:apend_B}. This function should tend toward a constant value at infinity, denoted by $M_{f(R)}^\infty$ as defined in \eqref{eq:M_inf} and depicted in Fig. \ref{fig:MfR}. The function $M_{f(R)}(r)$ oscillates around $M_{f(R)}(r)(1+U(r))$, the latter appearing as a band with a certain width. In our specific realization depicted in Figs. \ref{fig:Buch_comparation} and \ref{fig:MfR}, the mass that an observer at infinity would measure corresponds to $M_{f(R)}^\infty = 4.1\, {\rm km} = 2.78\,M_{\odot}$.

\begin{figure}
    \centering
    \includegraphics[width=0.35\textwidth]{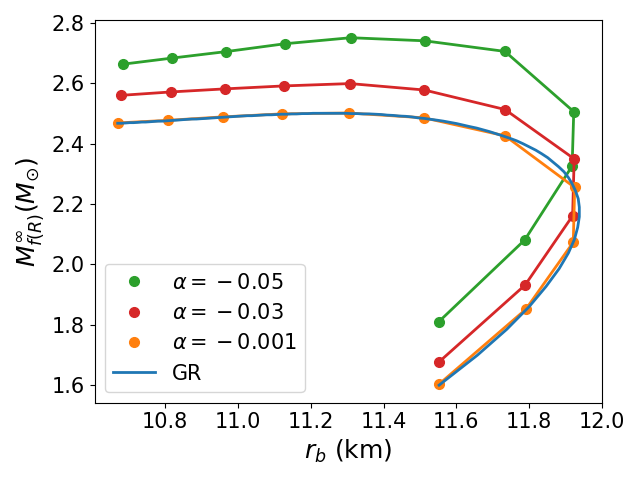}
    \caption{Mass-radius diagrams for GR and $f(R)=R+\alpha R^2$ model with $\alpha=-0.001,-0.03,-0.05\ {\rm km^2}$. We use the middle EOS and central pressures $p_c \in [1\cdot 10^{-4},1.5\cdot 10^{-3}]\,{\rm km}^{-2}$.}
    \label{fig:mass_radius}
\end{figure}

Once equipped with a method to calculate the mass measured by an observer at infinity for $f(R)$ theories, we can construct mass-radius diagrams by fixing the fluid EOS and considering different values for the central pressure $p_c$. This way, we can compare the mass-radius relation obtained in GR with the one obtained in the $f(R)=R+\alpha R^2$ models for different values of $\alpha$ following the method illustrated in Sec. \ref{Sec:IIC}. Thus, the mass-radius diagrams are shown in Fig. \ref{fig:mass_radius}. We can observe that, as the $f(R)$ theory deviates from GR, compact stars with higher asymptotic masses can exist. 
%This hints that in $f(R)$ gravity theories, compact objects might contain additional mass.
%
%

\section{Generalization of the Buchdahl limit in $f(R)$ theories}
\label{Sec:III}

\subsection{Derivation for general $f(R)$ models}
\label{Sec:IIIA}

To obtain a generalization of the Buchdahl limit for general $f(R)$ theories, we start with the following inequality 
\begin{eqnarray}\label{eq:result_appC}\nonumber
   && f_{R}(R(r_b)) c\left(r_b\right)-f_{R}\left(R(0)\right) c(0) \\ \label{eq:after_prop2}
   && \geq \dfrac{f_{R}(R(r_b))}{2} c(r_b) \left(\dfrac{1}{\sqrt{1-\dfrac{2m(r_b)}{r_b}}}-1\right) \,,
\end{eqnarray}
as obtained in Appendix \ref{App:apend_C}. On the other hand, using conditions \eqref{eq:cond_18} and \eqref{eq:cond_25}, we can state that
\begin{equation}
    f_{R}(R(r_b)) c\left(r_b\right)-f_{R}\left(R(0)\right) c(0) \leq f_{R}\left(R(0)\right) c(r_b).
\end{equation}
Hence, if we apply the above equation in Eq. \eqref{eq:result_appC} we get
\begin{equation}\label{eq:after_prop3}
    f_R(R(0)c(r_b) \geq \dfrac{f_{R}(R(r_b))}{2} c(r_b) \left(\dfrac{1}{\sqrt{1-\dfrac{2m(r_b)}{r_b}}}-1\right) \,,
\end{equation}
which with some further manipulations, becomes
\begin{equation}
    \sqrt{1-\dfrac{2m(r_b)}{r_b}}\left( 1 + 2 \dfrac{f_R(R(0))}{f_R(R(r_b))} \right) \geq 1\,.
    \label{Eq:43}
\end{equation}
Finally, by isolating the term $2m(r_b)$ we obtain the inequality
\begin{equation}\label{eq:Buch_fR1}
    2\,m(r_b) \leq\dfrac{4 \dfrac{f_{R}(R(0))}{f_{R}(R(r_b))}\left(1+\dfrac{f_{R}(R(0))}{f_{R}(R(r_b))}\right)}{\left(1+2 \dfrac{f_{R}(R(0))}{f_{R}(R(r_b))}\right)^2} \,r_b\,.
\end{equation}
As shown in Sec. \ref{Sec:IID}, the function
$m(r)$, according to its definition \eqref{eq:Buch_metric1} in terms of $A(r)$ and unlike the Schwarzschild case, typically oscillates outside the star.%around unity for sufficiently large distances. 
%
%Therefore, unlike the Schwarzschild case, the function $m(r)$ in the exterior possesses an oscillatory behavior
\footnote{
Nonetheless, $m(r)$ does correspond to the Misner-Sharp-Hern\'andez quasilocal mass, which in turn coincides with the Arnowitt-Deser-Misner mass at $r\rightarrow\infty$ for asymptotically flat spacetimes, as the ones considered herein.
}
Hence, to study the exterior spacetime,
the combination of \eqref{eq:Buch_metric1} and \eqref{eq:metric_functions}
allows us to do the following identification,
\begin{equation} \label{eq:exterior_m}
    m(r) = M_{f(R)}(r) - r\,V(r) %\dfrac{r_b\,U(r_b)}{2(1+U(r_b))}
    \,,
\end{equation}
%where $M_{f(R)}(r)$ is given by Eq. \eqref{eq:M_fR_def}.
%
%
where we have defined $V(r)=-U(r)/(2(1+U(r)))$.
Since at infinity, $A(r)$ and $B(r)$ should coincide, i.e., $U(r)$ must vanish there, the limits of $m(r)$ and $M_{f(R)}(r)$ at infinity should too. Thus, in order for the latter to be true, and given that the condition \eqref{eq:exterior_cond} needs to hold, one can conclude that
\begin{equation}
    \lim_{r\to\infty} rU(r) = 0\,,
\end{equation}
or in other words, $U(r)$ is a function that at large distances tends to zero faster than $1/r$. Due to the oscillatory behavior of $A(r)$ and $B(r)$, the value $U(r_b)$ could be either positive or negative depending on the chosen EOS, the $p_c$ values and the parameters of the $f(R)$ model under consideration.\footnote{In the case of the $f(R)$ Starobinsky model, the values of $\alpha$. For this choice, within the $p_c$ and $\alpha$ values considered throughout this investigation, and for the three EOS under consideration, the absolute value of the quotient between $r_bU(r_b)/2(1+U(r_b))$ and $M_{f(R)}(r_b)$ turns out to be lower than $5\cdot10^{-4}$. Thus, the last term of Eq. \eqref{eq:exterior_m} has a very small contribution. However, we could not get rid of this term since it can be negative.}
Consequently, using Eq. \eqref{eq:exterior_m} when evaluated at $r=r_b$, the inequality \eqref{eq:Buch_fR1} becomes
\begin{equation}\label{eq:Buch_MfR}
2 M_{f(R)}(r_b) \leq \dfrac{4 \dfrac{f_{R}(R(0))}{f_{R}(R(r_b))}\left(1+\dfrac{f_{R}(R(0))}{f_{R}(R(r_b))}\right) + V(r_b)}{\left(1+2 \dfrac{f_{R}(R(0))}{f_{R}(R(r_b))}\right)^2}\,r_b\,.
\end{equation}
As mentioned above, since the value of $M_{f(R)}(r_b)$ cannot be measured experimentally, our aim is to find a relationship between this quantity and the mass that an observer at infinity would actually measure, $M_{f(R)}^\infty$ as per \eqref{eq:M_inf}. Accordingly, we suggest the following relation,
\begin{equation}\label{eq:Buch_funcional0}
F[f(R(r_b))] M_{f(R)}^\infty \leq M_{f(R)}(r_b)\,,
\end{equation}
with $F[f(R(r_b))]$ a functional dependent on the function $f(R)$ evaluated at $r=r_b$. For $f(R)=R$ (GR), this functional must be unity, since in this case the mass at the edge of the star corresponds to the Schwarzschild mass measured by an observer at infinity. Additionally, for 
$f(R)\neq R$, due to the additional energetic content that $f(R)$ theories are able to host outside the star it happens that 
$M_{f(R)}^\infty > M_{f(R)}(r_b)$. Therefore, the functional $F$ should be smaller than 1 when considering deviations from GR. For all these reasons, we propose the expression
\begin{equation} \label{functional}
F[f(R(r_b))] = \left[1+\frac{f_{2R}(R(r_b))}{1{\rm km}^2}\right]^n\,,
\end{equation}
where $n$ is a positive real number and $f_{2R}$ satisfies \eqref{eq:cond_18}. Below we will test the validity of this ansatz for the Starobinsky quadratic models.
%This expression fully satisfies the conditions mentioned above.

\subsection{%The paradigmatic case of 
Starobinsky quadratic models}
\label{Sec:IIIB}

Let's now examine the validity of the functional \eqref{functional} for the case $f(R)=R+\alpha R^2$, in which
\begin{equation}
F[f(R(r_b))] = \left(1+2\dfrac{\alpha}{1\,{\rm km}^2}\right)^n\,.
\end{equation}
%which satisfies the mentioned conditions. In the above expression, we have written $\alpha/(1{\rm km}^2)$ to achieve a dimensionless expression, since in this model, $f_{2R}(R(r_b))=2\alpha$ has units of ${\rm km}^2$. 
%
If we redefine the parameter $\hat{\alpha} \equiv \alpha/(1{\rm km}^2)$, the relation \eqref{eq:Buch_funcional0} becomes
\begin{equation}\label{eq:Buch_funcional_n}
(1+2\hat{\alpha})^n M_{f(R)}^\infty \leq M_{f(R)}(r_b)\,.
\end{equation}
Based on the results obtained in Sec. \ref{Sec:IIC}, we calculate the relationship between $M_{f(R)}^\infty$ and $M_{f(R)}(r_b)$. This way, we can verify whether there exists a value of $n$ in \eqref{eq:Buch_funcional_n} for which the above inequality holds true for any of the three EOS considered throughout this investigation, for any value of $\alpha$ (provided that $|\hat{\alpha}|\ll 1$, i.e., small deviations from GR) and the range of central pressures leading to the existence of stable stars. As can be observed in Fig. \ref{fig:Buch1_} in Appendix \ref{App:apend_D}, it turns out that for $n=4$, the relation \eqref{eq:Buch_funcional_n} holds true in every possible case of interest. Further discussion about this choice is provided in Appendix \ref{App:apend_D}. Thus, we propose the relationship
\begin{equation}\label{eq:Buch_funcional}
\left(1+2\hat{\alpha}\right)^4 M_{f(R)}^\infty \leq M_{f(R)}(r_b)\,.
\end{equation}
%
%which only becomes an equality when $\hat{\alpha}=0$, as expected.
%
Finally, substituting \eqref{eq:Buch_funcional} into \eqref{eq:Buch_MfR}, we obtain 
\begin{equation}\label{eq:Buch_final}
    2\,M_{f(R)}^\infty \leq\dfrac{4 \dfrac{f_{R}(R(0))}{f_{R}(R(r_b))}\left(1+\dfrac{f_{R}(R(0))}{f_{R}(R(r_b))}\right) + V(r_b)}{(1+f_{2R}(R(r_b)))^4\left(1+2 \dfrac{f_{R}(R(0))}{f_{R}(R(r_b))}\right)^2}\;r_b\,,
\end{equation}
which can be understood as a generalized Buchdahl limit for $f(R)$ Starobinsky quadratic models and valid for the barotropic EOS whose pressure grows faster with density than in the soft case as explained in Appendix \ref{App:apend_D}. Analogous reasoning could be followed for either other competitive $f(R)$ models or barotropic EOS whose pressure grows slower with density than in the soft case. In any manner, expression \eqref{eq:Buch_MfR} remains valid for all $f(R)$ models for which conditions gathered in Secs. \ref{Sec:IIA} and \ref{Sec:IIB} and the end of the Appendix \ref{App:apend_C} hold,\footnote{The addition of thin shells or double layers at $r=r_b$ may modify the junction conditions \eqref{junction_conditions_fR} and consequently render the required procedure to find the asymptotic mass either different from the one explained here or unnecessary. Also, Eq. \eqref{eq:Buch_MfR} could then be different in those scenarios.} although a functional relation between the mass function evaluated at the edge of the star and the asymptotic mass---as Eq. \eqref{eq:Buch_funcional0}---would still be needed.

\subsection{Buchdahl limit in $f(R)$ theories embedded in a Schwarzschild exterior}
\label{Sec:IIIC}
In Sec. \ref{Sec:IIC}, we mentioned the well-known result stating that should the exterior spacetime be enforced to be Schwarzschild, not all necessary $f(R)$ junction conditions \eqref{junction_conditions_fR} would be in general satisfied. Thus, this scenario should be dismissed. However, for comparison purposes with the results of this investigation, and given the fact that the Schwarzschild metric is indeed a vacuum solution for wide classes of $f(R)$ theories, including the quadratic Starobinsky models, Eq. \eqref{eq:Buch_final} would become
\begin{equation}\label{eq:Buch_GR}
    2\,M \rvert_{\rm Schw} \leq \dfrac{4 \dfrac{f_{R}(R(0))}{f_{R}(0)}\left(1+\dfrac{f_{R}(R(0))}{f_{R}(0)}\right)}{\left(1+2 \dfrac{f_{R}(R(0))}{f_{R}(0)}\right)^2}\;r_b\,,
\end{equation}
where $M \rvert_{\rm Schw}$ denotes the obtained mass should the exterior have been imposed to be purely Schwarzschild, that is, $R(r\geq r_b)=0$. With the caveat mentioned above, the result in \eqref{eq:Buch_GR} would then be valid for a Schwarzschild exterior in any $f(R)$ theory, not only for the $f(R)=R+\alpha R^2$ model. Note that this relation is the same as in \cite{Goswami:2015dma}, since for a Schwarzschild exterior the additional term $V(r_b)$ in \eqref{eq:Buch_final} cancels out.

Using condition \eqref{eq:cond_18}, we can state that provided $R(0)<0$ then $f_{R}\left(R(0)\right)/f_{R}(0) \geq 1$.
When this quotient reaches unity, as for $\alpha = 0$, the usual GR Buchdahl limit $2M\leq(8/9)r_b$ is recovered. In the opposite scenario, i.e., should $f_{R}\left(R(0)\right)/f_{R}(0)$ be much larger than 1, the bound in \eqref{eq:Buch_GR} would tend to the usual Schwarzschild black hole limit $2M<r_b$. 
Consequently, as pointed out in \cite{Goswami:2015dma}, the assumption of a Schwarzschild exterior in $f(R)$ gravity allows us to accommodate stable and spherical stars whose mass lies in the $(8/9)r_b \leq 2M \leq r_b$ range, which is forbidden in GR. In the following, we shall elucidate what happens when Eq. \eqref{eq:Buch_final} is used, i.e., whenever all the $f(R)$ junction conditions are obeyed and therefore the exterior spacetime is not Schwarzschild.

\section{Observational consequences of the $f(R)$ Buchdahl limit}
\label{Sec:IV}

\subsection{Mass increment}
\label{Sec:IVA}
If we consider $f(R)$ terms as small perturbations around GR, that is, $|\hat{\alpha}|\ll 1$ in Starobinsky models, and develop the expression \eqref{eq:Buch_final} to second order in $\hat{\alpha}$, we obtain
\begin{equation}
\label{eq:mass_limit_general} 
    \dfrac{2M_{f(R)}^\infty}{r_b} <\dfrac{8}{9} \left[ 1 + \dfrac{V(r_b)}{8} + \hat{\alpha}\,C(r_b) + \hat{\alpha}^2\, D(r_b) \right] +\mathcal{O}(\hat{\alpha}^3)\,,  
\end{equation}
where 
\begin{eqnarray}
    C(r_b) &\equiv&  \dfrac{1}{3}\left(\hat{R}(0) - \hat{R}(r_b)\right)\left(1-V(r_b)\right) - V(r_b) - 8\,,\nonumber\\
    \\
    D(r_b) &\equiv& -\dfrac{2}{3}\left(\hat{R}(0)+4\right)\left(\hat{R}(0)-\hat{R}(r_b) \right) \left(1-V(r_b)\right) \nonumber\\
    &+& 10V(r_b) + 40\,,
\end{eqnarray}
and we have introduced the dimensionless Ricci scalar $\hat{R}\equiv R/\left(1\,{\rm km}^{-2}\right)$.

Then, we can define the relative mass increment as 
$\delta_M=(M_{f(R), max}^{\infty}-M_{\rm GR})/M_{\rm GR}$ 
where $M_{f(R), max}^{\infty}$ corresponds to the value that saturates the inequality in \eqref{eq:mass_limit_general} 
whereas $M_{\rm GR}=4/9\,r_b$ accounts for the mass saturating the usual GR Buchdahl limit. Thus, 
$\delta_M$ accounts for the relative mass difference---for the same EOS---as predicted by GR or quadratic Starobinsky $f(R)$ theories for a hypothetical same radius $r_b$. Thus
\begin{equation}
\label{eq:diferencia_masa_fR}
    \delta_{M}= \dfrac{V(r_b)}{8} + \hat{\alpha}\,C(r_b) + \hat{\alpha}^2\, D(r_b) + \mathcal{O}(\hat{\alpha}^3)\,.
\end{equation}
If an analogous procedure is followed departing from \eqref{eq:Buch_GR}, i.e., having naively assumed an Schwarzschild exterior, the Buchdahl limit there up to second order in $\hat{\alpha}$ provides
\begin{eqnarray}
    && \dfrac{2 M\rvert_{\rm Schw}}{r_b}<\dfrac{8}{9}\left[ 1 + \dfrac{1}{3}\hat{\alpha}\left(\hat{R}(0) - 24\right) - \right. \nonumber\\
    && \left. \dfrac{2}{3} \hat{\alpha}^2 \left( \hat{R}^2(0) + 4 \hat{R}(0)-60\right)\right]+\mathcal{O}(\hat{\alpha}^3)\,.
\end{eqnarray}
Thus, in this case the mass increment would satisfy
% is given by
%
\begin{eqnarray}\label{eq:diferencia_masa_goswami}
    &&\delta_{M}\rvert_{\rm Schw}=\dfrac{1}{3}\hat{\alpha}\left(\hat{R}(0) - 24\right) - \nonumber\\
    &&\dfrac{2}{3} \hat{\alpha}^2 \left( \hat{R}^2(0) + 4 \hat{R}(0)-60\right) +\mathcal{O}(\hat{\alpha}^3) \,.
\end{eqnarray}
%where we can identify $C(r_b)=\frac{1}{3}(\hat{R}(0) - 24)$ and $D(r_b)=-\frac{2}{3} ( \hat{R}^2(0) + 4 \hat{R}(0)-60)$.

\subsection{Results}
\label{Sec:IVB}
Next, we will extract some consequences of using either our generalized expression \eqref{eq:Buch_final} or the usual GR Buchdahl limit. This way, the ability of \eqref{eq:Buch_final} to constrain the permitted mass-radius interval for $f(R)$ relativistic stars would be elucidated. Once again, the considered EOS in the following is the middle one for the reasons explained in Appendix \ref{App:apend_D}.
\subsubsection{Mass upper limits}
\begin{figure}
\centering
\subfigure[$\alpha=-0.002\,\text{km}^2$]{
\label{subfig:5a}
\includegraphics[width=0.23\textwidth]{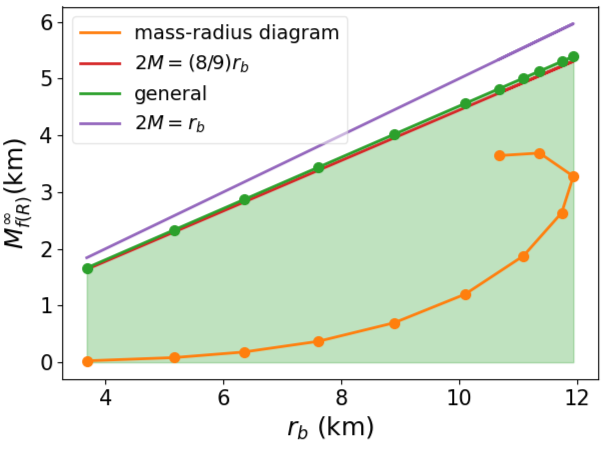}}
\subfigure[$\alpha=-0.01\,\text{km}^2$]{
\label{subfig:5b}
\includegraphics[width=0.23\textwidth]{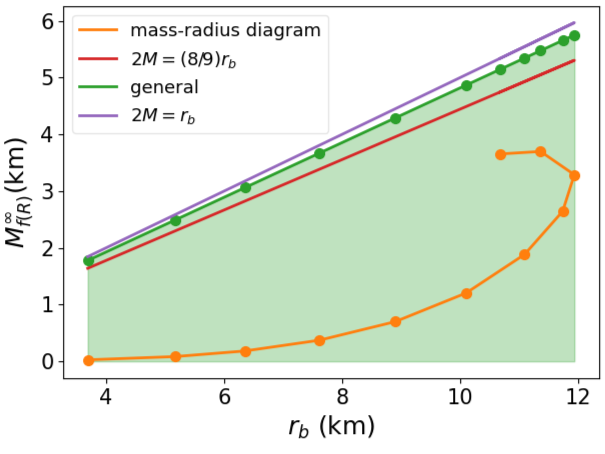}}
\subfigure[$\alpha=-0.015\,\text{km}^2$]{
\label{subfig:5c}
\includegraphics[width=0.23\textwidth]{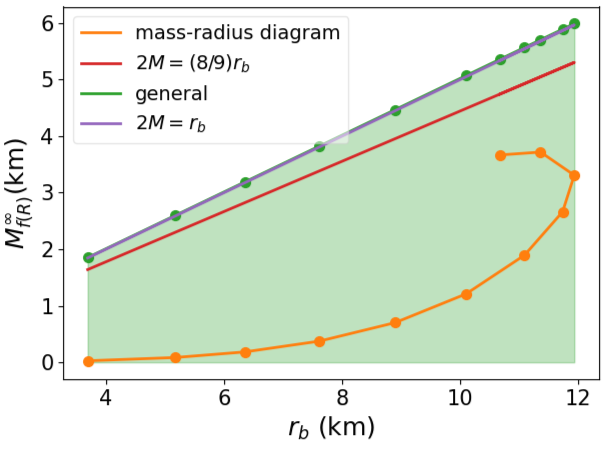}}
\subfigure[$\alpha=-0.02\,\text{km}^2$]{
\label{subfig:5d}
\includegraphics[width=0.23\textwidth]{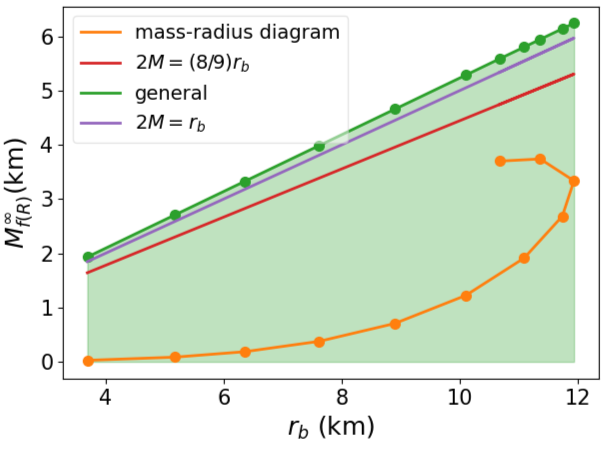}}
\caption{Comparison of Buchdahl limit as per Eq. \eqref{eq:Buch_final} (``general'') with the one obtained in GR $[2M=(8/9)r_b]$ and the Schwarzschild black hole limit ($2M=r_b$), for the $f(R)=R+\alpha R^2$ model with (a) $\alpha=-0.002$, (b) $\alpha=-0.01$, (c) $\alpha=-0.015$, (d) $\alpha=-0.02\,\text{km}^2$. The green dots represent the performed simulations. We have considered the middle EOS and central pressures $p_c \in [1\cdot 10^{-6},1.5\cdot 10^{-3}]\,{\rm km}^{-2}$. We also depict the mass-radius diagram for these values of $\alpha$ and $p_c$ interval. 
%Smaller values of $\alpha$ correspond to greater deviations from GR.
}
\label{fig:Buch2_comparation}
\end{figure}

Thus, in Fig. \ref{fig:Buch2_comparation} we depict the mass, which an observer at infinity would measure, saturating inequality \eqref{eq:Buch_final} as a function of the star's radius. We notice that as the parameter $\alpha$ approaches zero, our generalized Buchdahl limit converges toward the corresponding GR Buchdahl limit, the former always being above the latter. Additionally, as the value of $\alpha$ takes higher negative values, there exists a more noticeable discrepancy between our limit and that of GR. Also, our generalized upper bound \eqref{eq:Buch_final} %approaches the GR Schwarzschild BH limit $2M=r_b$. This trend 
results in that for values of $\alpha$ lower than $-0.015\, \text{km}^2$, the GR  Schwarzschild BH limit can be exceeded.
This does not mean that our numerical solutions would correspond to black holes, since $2M=r_b$ would be the position of the event horizon in GR on a purely Schwarzschild exterior. However, the latter will not be the horizon position for calculations performed in the context of $f(R)$ theories, whose herein solutions do not present horizons since they correspond to regular stars.
In fact, we see that the radius of stars in $f(R)$ theories can be smaller than twice its mass $M_{f(R)}^{\infty}$. Consequently, an object forced to be a black hole in GR, could be a stable star in $f(R)$ theories, see for instance Fig. \ref{fig:Buch_comparation} or Fig. \ref{subfig:5d}. 
Finally, in Fig. \ref{fig:Buch2_comparation} we can observe how for the displayed $f(R)$ realizations, the actual mass-radius diagrams satisfy the three limits mentioned therein. Because for the other two EOS no significant changes are observed when performing such analyses, we decided not to present such results. Thus, our conclusions hold valid for the three EOS under study.\\

\subsubsection{Mass increment} 
\begin{figure}
\centering
\includegraphics[width=0.35\textwidth]{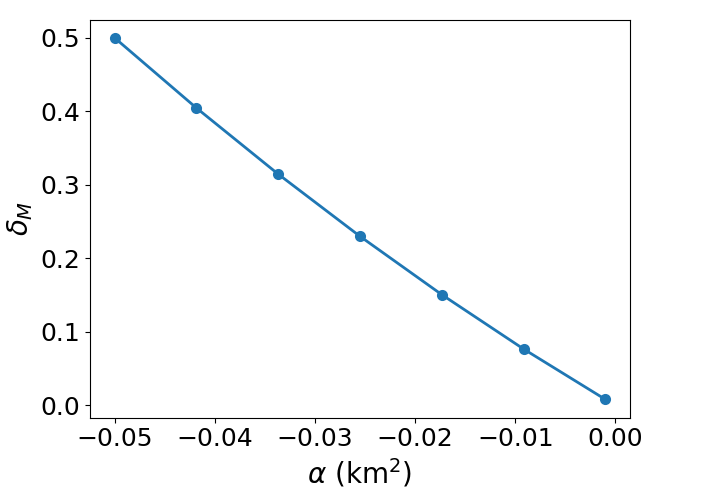}
\caption{
Relative mass increment introduced by $f(R)=R+\alpha R^2$ model for different values of $\alpha$. For illustrative purposes, we have considered the middle EOS and a central pressure $p_c=5\cdot 10^{-4} \text{km}^{-2}$. The dots represent the seven simulations performed.
This curve shows the result for the (general) Buchdahl limit as obtained using Eq. 
\eqref{eq:diferencia_masa_fR} up to second order in $\alpha$.
}. 
\label{fig:Buch3_comparation}
\end{figure}

In order to quantify the change in the mass upper limits as provided by either the usual GR prediction or the obtained Buchdahl limit expression \eqref{eq:Buch_final}, in Fig. \ref{fig:Buch3_comparation} we depict the relative mass increment as calculated through \eqref{eq:diferencia_masa_fR}. Furthermore, this figure also displays the relative mass increment should a Schwarzschild exterior be assumed, i.e., as calculated through the expression 
% the mass increment corresponding to the result  obtained in \cite{Goswami:2015dma}, which we calculated using equation 
%
\eqref{eq:diferencia_masa_goswami}.
In both cases, we observe that the relative mass differences are positive and increase as we deviate from GR, that is, as the value of $\alpha$ is more negative. Complementarily, we note that the additional mass introduced by virtue of a Schwarzschild exterior, within the range of $\alpha$ values under consideration, lies below a $0.01\%$ increment in the studied range of $\alpha$, making it very challenging to measure. However, the additional mass introduced by our correctly joined $f(R)$ result is of the order of $10\%$, for these range of values of $\alpha$. Therefore, we conclude that the use of a correctly matched, that is, non-Schwarzschild, exterior would lead to a more appreciable mass difference when compared to GR predictions.\\

\begin{figure}
\centering
\includegraphics[width=0.35\textwidth]{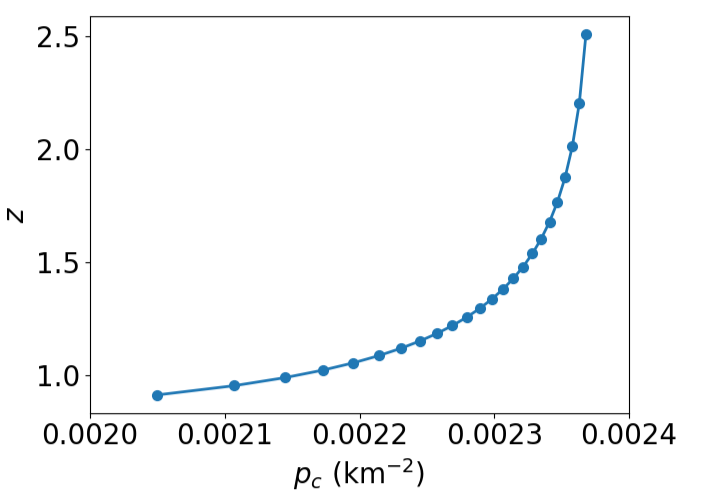}
\caption{Gravitational redshift for the $f(R)=R+\alpha R^2$ model with $\alpha=-0.05 \,\text{km}^2$ and middle EOS, for different values of $p_c$. 
%Smaller values of $\alpha$ correspond to greater values of z.
}
\label{fig:z_comparation}
\end{figure}

\subsubsection{Gravitational redshift} 
For static spacetimes of the form of \eqref{eq:metric}, the surface gravitational redshift is given by $z=\sqrt{B^{-1}(r_b)} - 1$. Since in the context of $f(R)$ theories there is no analytical expression for $B(r_b)$, it is not possible to establish an analytical bound for this quantity and therefore, a maximum value of $z$ cannot be foretold in this context. 
Instead, in Fig. \ref{fig:z_comparation} we depict the gravitational redshift for different values of $p_c$ for a fixed $\alpha$ and middle EOS. As we can see, the greater the values of $p_c$, the higher the values of the gravitational redshift. For $p_c \gtrsim 0.00235 \, {\rm km}^{-2}$
the value of $z$ exceeds 2, i.e., it surpasses the GR limit, whereas for $p_c>0.0024 \, {\rm km}^{-2}$ there are no static stars. Analogous conclusion is obtained for other values of $\alpha$ and the two other EOS considered throughout this communication, for which no significant changes are observed, so we have not depicted the corresponding figures. Hence, we conclude that in $f(R)$ models it is possible that compact stars may provide gravitational redshift values higher than the bound imposed by the usual GR Buchdahl limit.

\section{Conclusions}
\label{Sec:V}
In this investigation we have set bounds on the mass-radius relation for static and spherically symmetric neutron stars endowed with realistic barotropic equations of state when embedded in $f(R)$ spacetimes.
The fact that for such gravitational theories the junction conditions are more restrictive than their Einsteinian counterparts results in the exteriors being distinct from the Schwarzschild spacetime, i.e., the Ricci scalar in the exterior is not identically zero. Although the process throughout this communication is fully general, this fact has forced us to obtain, first, a general Buchdahl limit---see Eq. \eqref{eq:Buch_MfR}---valid for general viable $f(R)$ theories and, second, for the class of paradigmatic Starobinsky quadratic $f(R)$ models, see Eq. \eqref{eq:Buch_final}. 

In this context, after having assumed very generic conditions of regularity and thermodynamic stability in the interior of the star and imposed all the junction conditions at the stellar surface to obtain asymptotically flat---although not purely Schwarzschild---exterior, we transparently demonstrated that the mass-to-radius ratio of the star is bounded from above, and this is a stricter bound than when the Schwarzschild exterior is assumed \textit{ad hoc}. In other words, we generalized the Buchdahl bound on static stars in $f(R)$ theories.
We also showed that, whenever $f(R) \neq R$, this upper bound is less restrictive than the usual Einsteinian Buchdahl limit.  
Hence, in principle, we can pack additional effective mass in a stable compact star in these theories. These extra-massive stars could potentially provide a solution, among others, to the dark matter problem ({\it cf.} \cite{Cembranos:2008gj,Nojiri:2008nt}), manifesting through astrophysical observations of compact objects, such as the gravitational redshift. As widely known, objects in general relativity with a gravitational redshift greater than 2 are typically unstable and in the process of collapsing into a black hole. The discovery of stable stars with $z>2$ would suggest either the existence of an exotic, previously unknown type of object or indicate that general relativity itself is an incomplete theory in strong-gravity regimes which needs to be modified or replaced by a more comprehensive theory. One such alternative is $f(R)$ gravity, where the Buchdahl limit has been shown here to be altered, allowing the existence of stable, spherically symmetric stars with a gravitational redshift exceeding 2. Once this fact has been fully determined herein, natural steps to follow would be revisiting the stability criteria for compact objects, as well as both the description of gravitational collapse and the formation of event horizons. All these phenomena may potentially offer deeper insights into the nature of strong gravitational fields.

Finally, since we have only verified the validity of the inequality \eqref{eq:Buch_final} for the family of functions $f(R)=R+\alpha R^2$, as well as the neutron-matter equations of state described above, we can only ensure that our conclusions are valid for this paradigmatic family of $f(R)$ models, which may indeed encapsulate corrections dominant in strong-gravity regimes. Further investigation for other competitive classes of $f(R)$ models, as well as equations of state in the LIGO-Virgo-KAGRA permitted windows ({\it cf.} \cite{LIGOScientific:2018cki, Dietrich:2020efo}), is in progress.

\section*{Acknowledgements}
%
%{\bf Acknowledgments:} 
The authors would like to thank Jose Gonz\'{a}lvez P\'{a}rraga for insightful comments and Adri\'an Casado-Turri\'on for the comprehensive reading of the manuscript and suggestion of bibliographic references.
AdlCD acknowledges support from BG20/00236 action (MCINU, Spain), NRF Grant No. CSUR23042798041, CSIC Grant No. COOPB23096, Project SA097P24 funded by Junta de Castilla y Le\'on (Spain), and Grant No. PID2021-122938NB-I00 funded by MCIN/AEI/10.13039/501100011033 and by ERDF ``A way of making Europe''.

\section{Appendixes}
\label{Sec:Appendices}
%
%The following Appendix \ref{App:apend_A} includes the derivation of the pertinent equations of motion and a detailed explanation of the chosen initial conditions. Then, Appendix \ref{App:apend_B} displays  %short discussion of the method to find asymptotically flat solutions. Next, in Appendix \ref{App:apend_C} we present the formalism we use to obtain the expression for the Buchdahl limit \eqref{eq:Buch_MfR}. Finally, Appendix \ref{App:apend_D} provides some crosschecks to guarantee the upper-bound ansatz \eqref{eq:Buch_funcional} made for quadratic Starobinsky $f(R)$ models, which relates the asymptotic mass and the mass function when evaluated at the radius of the star.
%

\appendix
\section{Equations of motion and initial conditions}
\label{App:apend_A}
\renewcommand{\theequation}{A.\arabic{equation}}

Since our metric is static, we consider the matter fluid to be at rest, thus the spatial components of velocity being zero. Furthermore, a normalized velocity ($u_\mu u^\mu = -1$) leads to the components of the four-velocity being $u = \sqrt{B}\left( 1,0,0,0 \right)$. This renders the matter energy-momentum tensor diagonal
as in \eqref{eq:perfectfluid}.

Next, the components of the Ricci tensor are derived from the Christoffel symbols as
\begin{equation}
R_{\mu \nu}=R_{\mu \sigma \nu}^\sigma=\partial_\sigma \Gamma_{\mu \nu}^\sigma-\partial_\nu \Gamma_{\mu \sigma}^\sigma+\Gamma_{\sigma \rho}^\sigma \Gamma_{\mu \nu}^\rho-\Gamma_{\nu \rho}^\sigma \Gamma_{\mu \sigma}^\rho \,.
\end{equation}
For this metric, the nonzero Christoffel symbols are
\begin{align}\nonumber
& \Gamma_{r r}^r=\dfrac{A'}{2 A}, \quad \Gamma_{t t}^r=\dfrac{B'}{2 A}, \quad \Gamma_{\phi \phi}^r=-\dfrac{r \sin ^2 \theta}{A}\,,  \\ \nonumber
& \Gamma_{\theta \theta}^r=-\dfrac{r}{A}, \quad \Gamma_{\theta r}^\theta=\Gamma_{\phi r}^\phi=\dfrac{1}{r}, \quad \Gamma_{t r}^t=\dfrac{B'}{2 B}\,, \\
& \Gamma_{\phi \phi}^\theta=-\sin \theta \cos \theta, \quad \Gamma_{\phi \theta}^\phi=\cot \theta\,,
\end{align}
which imply that the only nonzero components of the Ricci tensor are
\begin{gather}
R_{t t}=\dfrac{B''}{2 A}-\dfrac{B'}{4 A}\left(\dfrac{B'}{B}+\dfrac{A'}{A}\right)+\dfrac{B'}{r A}, \\
R_{r r}=-\dfrac{B''}{2 B}+\dfrac{B'}{4 B}\left(\dfrac{B'}{B}+\dfrac{A'}{A}\right)+\dfrac{A'}{r A}, \\
R_{\theta \theta}=1-\dfrac{1}{A}-\dfrac{r}{2 A}\left(\dfrac{B'}{B}-\dfrac{A'}{A}\right)\,, \\
R_{\phi \phi}=\sin^2\theta R_{\theta \theta}.
\end{gather}
Hence, the Ricci scalar $R \equiv g^{\mu\nu}R_{\mu\nu}$ yields
\begin{gather}\nonumber
R = \dfrac{B'}{2 A B}\left(\dfrac{A'}{A}+\dfrac{B'}{B}\right)-\dfrac{B''}{A B} \\ \label{eq:R}
-\dfrac{2 B'}{r A B}+\dfrac{2 A'}{r A^2}-\dfrac{2}{A r^2}+\dfrac{2}{r^2}\,.
\end{gather}
Once we know the components of the tensors $g_{\mu\nu}$, $T^M_{\mu\nu}$, and $R_{\mu\nu}$, their substitution into the field equations \eqref{eq:field_ecs} result in three independent equations,
\begin{gather}\nonumber
\dfrac{B''}{2 A}-\dfrac{B'}{4 A}\left(\dfrac{B'}{B}+\dfrac{A'}{A}\right)+\dfrac{B'}{r A} =\dfrac{1}{f_R}\left[\kappa \rho B \right. \\ \label{eq:1edo0}
\left. - B\left(\dfrac{A'}{2 A^2}-\dfrac{2}{r A}\right) f_R'+\dfrac{B}{A} f_R''-\dfrac{B}{2}f(R)\right]\,,
\end{gather}
\begin{gather}\nonumber
-\dfrac{B''}{2 B}+\dfrac{B'}{4 B}\left(\dfrac{B'}{B}+\dfrac{A'}{A}\right)+\dfrac{A'}{r A} =\dfrac{1}{f_R}\left[\kappa p A \right. \\ \label{eq:2edo0}
\left.-\left(\dfrac{B'}{2 B}+\dfrac{2}{r}\right) f_R'+\dfrac{A}{2}f(R)\right]\,,
\end{gather}
\begin{gather}\nonumber
1-\dfrac{1}{A}-\dfrac{r}{2 A}\left(\dfrac{B'}{B}-\dfrac{A'}{A}\right) = \dfrac{1}{f_R}\left[\kappa p r^2 -\dfrac{r^2}{A} f_R'' \right. \\ \label{eq:3edo0}
\left. -\left(\dfrac{B' r^2}{2 A B}-\dfrac{A' r^2}{2 A^2}+\dfrac{r}{A}\right) f_R'+\dfrac{r^2}{2}f(R)\right]\,.
\end{gather}
Finally, by imposing the conservation of the matter energy-momentum tensor, i.e., $\nabla^\mu T^M_{\mu\nu}=0$, we obtain \eqref{eq:4edo}
%\begin{equation}
%p' = -\dfrac{p+\rho}{2}\dfrac{B'}{B}\,,
%\end{equation}
which completes our system of four independent equations. 

Now, to enable numerical solving of the equations, our goal is to isolate the higher-order derivatives and express them in terms of lower derivatives the other functions. The combination $\dfrac{\eqref{eq:1edo0}}{2B}+\dfrac{\eqref{eq:2edo0}}{2A}+\dfrac{\eqref{eq:3edo0}}{r^2}$ yields
\begin{gather}\nonumber
    -\dfrac{1}{A r^2}+\dfrac{A'}{r A^2}+\dfrac{1}{r^2}=\dfrac{1}{f_R}\left[\dfrac{\kappa}{2}(\rho+3 p)-\dfrac{1}{2 A} f_R^{\prime \prime} \right.\\ \label{eq:4edo0}
    \left. -\left(\dfrac{3 B'}{4 A B}-\dfrac{A'}{4 A^2}+\dfrac{1}{r A}\right) f_R'+\dfrac{1}{2} f(R)\right] \,.
\end{gather}
To isolate $A'$, we need to express $f_R''$ in terms of lower-order derivatives. To achieve this, by combining the expressions $\dfrac{3A}{r^2}$ from Eq. \eqref{eq:3edo0}, $\dfrac{A}{B}$ from Eq. \eqref{eq:1edo0}, and $-2A$ from Eq. \eqref{eq:4edo0}, 
%and solving for $f_R''$, 
we obtain the expression
\begin{align}\nonumber
& f_R\left(\dfrac{A R}{2}-\dfrac{A'}{2 r A}-\dfrac{2 A}{r^2}+\dfrac{2}{r^2}+\dfrac{3 B'}{2 r B}\right)+f_R'\left(\dfrac{A'}{2 A}+\dfrac{1}{r}\right)\\
& = f_R^{\prime \prime}\, .
\end{align}
By introducing this equation into \eqref{eq:4edo0} and rearranging the terms, %\footnote{We use the relation $f_R'=f_{2 R} R'$.}, 
we obtain Eq. \eqref{eq:1edo}. On the other hand, since $f_R^{\prime \prime}=f_{3 R} R^{\prime 2}+f_{2 R} R^{\prime \prime}$ and considering the expression \eqref{eq:R} for the Ricci scalar, we can isolate the term $R''$, resulting in Eq. \eqref{eq:3edo}. Finally, the expression for $B''$ is obtained from Eq. \eqref{eq:2edo0}, where $A'$ is substituted with the relation given in Eq. \eqref{eq:1edo}.
With this, the system of four independent differential equations \eqref{eq:1edo}--\eqref{eq:4edo}, in which the higher-order derivatives are isolated on the left-hand sides, is obtained.

\section{Shooting method to obtain $B(0)$}\label{App:apend_B}
\begin{figure}
    \centering
    \includegraphics[width=0.35\textwidth]{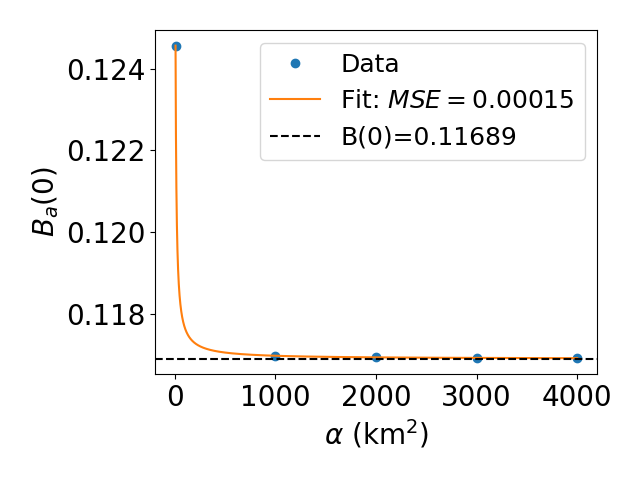}
    \caption{Initial value $B_a(0)$ as a function of the parameter $a$ that satisfies the boundary condition \eqref{eq:appendB_cond2}. When $a\to\infty$, we obtain $B_\infty(0)=0.11689$. We have used $\alpha=-0.05 \text{km}^2$, the middle EOS and a central pressure $p_c=5\cdot 10^{-4}{\rm km}^{-2}$.}
    \label{fig:shooting_Balpha}
\end{figure}

\renewcommand{\theequation}{B.\arabic{equation}}
Herein we shall follow the procedure outlined in \cite{AparicioResco:2016xcm}, which is valid for any $f(R)$ theory to determine the value of $B(0)$. Such a procedure solely relies on the condition
\begin{equation}\label{eq:appendB_cond1}
\lim_{r \to \infty} B(r) = 1\,,
\end{equation}
that is, at very large distances the metric should recover the Minkowskian limit. 
Whence the main idea behind the shooting method is to start with an arbitrary value of $B(0)$ and integrate the system of equations up to a certain radial distance. 
To apply this method, we need to choose a sufficiently large reference-radial value up to which the integration is performed. Let us denote this radial distance as $a r_b$, where $r_b$ is the radius of the star and $a>1$ is a dimensionless parameter. For each value of $a$, we choose $B_a(0)$ such that
\begin{equation}\label{eq:appendB_cond2}
B_a(a r_b)=1.
\end{equation}
We then increase the value of $a$ and calculate the corresponding $B_a(0)$ satisfying \eqref{eq:appendB_cond2}. Using this, we can plot the obtained values of $B_a(0)$ as a function of $a$ and perform a fitting using a function of the form $b_1+b_2/a^{b_3}$, where the $b^i$ are fitting parameters.\footnote{We use a nonlinear least-squares method to fit the function to data. The error of the fit can be calculated using errors propagation with the fitting function and using the error of each $b_{i=1,2,3}$.} An example of this fitting is shown in Fig. \ref{fig:shooting_Balpha} for the case of $f(R)=R+\alpha R^2$ with $\alpha=-0.05 \,{\rm km}^2$ and a central pressure $p_c=5\cdot 10^{-3}\,{\rm km}^{-2}$. 

Then by using this fit, we can calculate $B_{a\to\infty}(r=0)$, which corresponds to the initial value of $B(0)$ satisfying condition \eqref{eq:appendB_cond1}. With this, we can then solve the system of differential equations to obtain the functions $A(r)$, $B(r)$, $R(r)$, and $p(r)$ guaranteeing that all the initial, junction and asymptotical---among them \eqref{eq:appendB_cond2}---conditions are satisfied.

\section{Formalism to obtain the Buchdahl limit in $f(R)$ theories}\label{App:apend_C}
\renewcommand{\theequation}{C.\arabic{equation}}

%We use a similar reasoning as explained in \cite{Goswami:2015dma} to obtain an expression for the modified Buchdahl limit for any exterior, not necessarily Schwarzschild.

Provided conditions \eqref{eq:cond_18}--\eqref{eq:cond_26} are satisfied, the following reasoning leading to the inequality given in \eqref{eq:Buch_MfR} would be valid for any static and spherically symmetric exterior, not necessarily Schwarzschild, in the framework of metric $f(R)$ theories. In such a general context, although the matter pressure indeed cancels at the star radius as per \eqref{matter_rho_P_at_rb}, %i.e., $p\left(r_b\right)=0$, 
the total $p_r$ component defined in Eq. \eqref{eq:rho_p_(r)} does not necessarily vanish at the star radius for general $f(R)\neq R$.
In fact, using Eq. \eqref{eq:energy_mom_curvature}, we have
\begin{equation}
    p_r^R = (T^R)^r_r= \dfrac{1}{\kappa f_R}\left[\dfrac{1}{2}(f-Rf_R) + \nabla_r\nabla^r f_R - \square f_R  \right]\,.
\end{equation}
%
%which is identically zero in GR. 
Now, evaluating Eq. \eqref{eq:Buch_pr} at $r=r_b$ and isolating for $c'(r_b)$, we obtain
\begin{equation}
\label{eq:c_prime_rb}    c'\left(r_b\right)=\dfrac{m(r_b)}{r_b^2} \dfrac{c(r_b)}{1-\dfrac{2m(r_b)}{r_b}} + \dfrac{p_r^R(r_b) r_b}{2}\dfrac{c(r_b)}{1-\dfrac{2m(r_b)}{r_b}}\,.
\end{equation}
Accordingly, we shall prove the following proposition:\\ 

{\it Proposition 1.} In the general case satisfying conditions \eqref{eq:cond_18}--\eqref{eq:cond_26}, it holds that for any radius $r$ in the interior spacetime, the function $f_{R}(R(r)) c'(r)$ is bounded from below.\\

{\it Proof.} Using the condition \eqref{eq:cond_26} on Eq. \eqref{eq:cond_17} we get
\begin{eqnarray}    \label{first_prop_1}
&&f_{R}(r) \dfrac{{\rm d}}{{\rm d} r}\left(\frac{c'(r)}{r}\sqrt{1-\dfrac{2m(r)}{r}} \right)
\nonumber\\
&&\,+\,c(r) \dfrac{{\rm d}}{{\rm d} r}\left(\frac{f_R'(r)}{r}\sqrt{1-\dfrac{2m(r)}{r}}\right) \leq 0\,.
    \end{eqnarray}
Thus, \eqref{first_prop_1} can be integrated from $r$ (with $r\in [0,r_b]$) to $r_b$ and reordered, yielding 
\begin{eqnarray}
&&\left[\frac{c' f_R}{r}\sqrt{1-\dfrac{2m(r)}{r}} \right]_r^{r_b}  +\left[\frac{c f_R'}{r}\sqrt{1-\dfrac{2m(r)}{r}} \right]_r^{r_b}\nonumber\\
&&- 2 \int_r^{r_b} 
\frac{c'(\tilde{r}) f_{2R}(R(\tilde{r})) R'(\tilde{r}) }{\tilde{r}}\sqrt{1-\dfrac{2m(\tilde{r})}{\tilde{r}}}
{\rm d}\tilde{r} \leq 0\,.\nonumber\\
&&
\label{C4_eqn}
\end{eqnarray}
By reordering \eqref{C4_eqn} and using \eqref{eq:c_prime_rb} we obtain that $\forall r\in [0,r_b]$,
\begin{eqnarray}
%\label{Proposition1_result}]
\label{eq:prop3_demo}
&& \dfrac{{\rm d}}{{\rm d} r}\left( f_{R}(R(r)) c(r) \right) \geq 
\dfrac{c(r_b)\,r}{\sqrt{1-\dfrac{2 m(r_b)}{r_b}}\sqrt{1-\dfrac{2 m(r)}{r}}}\\
&&\times
\left[
f_R(R(r_b))\left(\dfrac{m(r_b)}{r_b^3}+\dfrac{p^R_r(r_b)}{2}\right)
\right.\nonumber\\
&&\left.+\,\dfrac{f_{2R}(R(r_b))R'(r_b)}{r_b}\left(1-\dfrac{2m(r_b)}{r_b}\right)\right] \, \nonumber\\
&&-\, \dfrac{2r}{\sqrt{1-\dfrac{2 m(r)}{r}}} \int_r^{r_b}\frac{c'(\tilde r) f_{2R}(R(\tilde r)) R'(\tilde r)}{\tilde r}\sqrt{1-\dfrac{2m(\tilde r)}{\tilde r}}
{\rm d}\tilde{r} \nonumber \,. 
\end{eqnarray}  
\begin{flushright}
$\blacksquare$
\end{flushright}
To understand the behavior of the terms that appear on the right-hand side of \eqref{eq:prop3_demo}, we need to consider a specific $f(R)$ model. For example, we analyze the case $f(R)=R+\alpha R^2$, for which we have obtained the solution of the field equations in Sec. \ref{Sec:IIA}. For all the range of studied values of the parameter $\alpha$ and central pressures we have checked that the conditions
\begin{equation}
\label{eq:hipotesis1}
    p_r^R(r_b)>0\quad \text{and} \quad R'(r_b)<0
\end{equation}
are satisfied. 
Therefore, in Eq. \eqref{eq:prop3_demo} we could drop the term containing $p_r^R$, as well as the term proportional to $R'(r_b)$ without invalidating the inequality, since both have a positive contribution. 

However, the sign of the last term in \eqref{eq:prop3_demo}, which includes an integral, depends on the value of $r$, i.e., on the lower limit in the integral expression. In fact, for most of the values of $r$, the mentioned term turns out to be positive. However, for values of $r$ near to the radius of the star, $R'(r)$ is usually negative,\footnote{At least in the quadratic $f(R)$ realizations in a wide interval of parameter $\alpha$ and central pressures we have performed.} and so does the last term of \eqref{eq:prop3_demo}, so we cannot in principle drop it from the inequality. For this reason, we numerically compared the terms involving $p_r^R$ and $R'(r_b)$ in \eqref{eq:prop3_demo} against this last term. By doing so, we have always obtained a positive contribution. Therefore, we can drop together these three terms in \eqref{eq:prop3_demo}, so the next inequality holds true
\begin{equation}
\label{eq:Proposition1_result_tris}
 \dfrac{{\rm d}}{{\rm d} r}\left( f_{R}(R(r)) c(r) \right) \geq \dfrac{f_{R}(R(r_b))}{r_b^3} \dfrac{m(r_b) c(r_b) r}{\sqrt{1-\dfrac{2 m(r_b)}{r_b}}\sqrt{1-\dfrac{2 m(r)}{r}}}\,.
\end{equation} 
Since the last expression holds $\forall r\in [0,r_b]$ and the term on the right-hand side is positive, we can integrate \eqref{eq:Proposition1_result_tris} in such an interval without changing the inequality. Thus, 
\begin{eqnarray}
\label{Proposition1_result_cuatris}
\nonumber
&&f_{R}(R(r_b)) c\left(r_b\right)-f_{R}\left(R(0)\right) c(0) \\ 
&& \geq \dfrac{f_{R}(R(r_b))}{r_b^3} \dfrac{m(r_b) c(r_b)}{\sqrt{1-\dfrac{2 m(r_b)}{r_b}}} \int_0^{r_b} \dfrac{r\,{\rm d} r}{\sqrt{1-\dfrac{2 m}{r}}} \,.
\end{eqnarray}
At this stage, in order to simplify the integral term on the right-hand side of the last equation, 
we use the following proposition.\\

{\it Proposition 2.} In general, for any stellar model that is thermodynamically stable and satisfies conditions \eqref{eq:cond_18}--\eqref{eq:cond_26}, the integral of $r\left(1 -\dfrac{2m(r)}{r}\right)^{-1/2}$ between 0 and $r_b$ is bounded from below. \\

{\it Proof.} Using condition \eqref{eq:cond_26}, in the interior of the star we obtain the condition
\begin{equation}
\label{eq:cond_prop2}
    2 m(r_b) \dfrac{r^3}{r_b^3} \leq 2 m(r) \quad \forall r\in [0,r_b]\,.
\end{equation}
Therefore, it is straightforward to show
\begin{eqnarray}
\label{result_Prop2}
    \nonumber
        \int_0^{r_b} \dfrac{r {\rm d} r}{\sqrt{1-\dfrac{2 m(r)}{r}}} \geq \int_0^{r_b} \dfrac{r {\rm d} r}{\sqrt{1-\dfrac{2 m(r_b) r^2}{r_b^3}}} \\
        = \dfrac{r_b^3}{2 m(r_b)}\left(1- \sqrt{1-\dfrac{2m(r_b)}{r_b}}\right)\,.
    \end{eqnarray}    
\begin{flushright}
$\blacksquare$
\end{flushright}
Finally, the use of \eqref{result_Prop2} in \eqref{Proposition1_result_cuatris} leads to the expression \eqref{eq:result_appC} given in Sec. \ref{Sec:IIIA}.

To summarize the conditions used since \eqref{eq:prop3_demo}: first, we assume $p_r^R(r_b)>0$ and $R'(r_b)<0$, consistent with our results for the field equations. Next, we note that the terms involving $p_r^R$ and $R'(r_b)$ are always positive and exceed the integral term in \eqref{eq:prop3_demo}. This allows us to drop the three terms from the inequality, as the hole contribution remains positive. Using the properties of $f_R$, $m$ and $c$, we establish that the right-hand side of \eqref{eq:Proposition1_result_tris} is positive. Consequently, both sides of the expression can be integrated without changing the inequality's sign. Finally, the expression \eqref{eq:cond_prop2} follows directly from the condition \eqref{eq:cond_26}.

\section{Validity of relation \eqref{eq:Buch_funcional}}
\label{App:apend_D}

\begin{figure}
    \centering
    \subfigure[]{
        \includegraphics[width=0.23\textwidth]{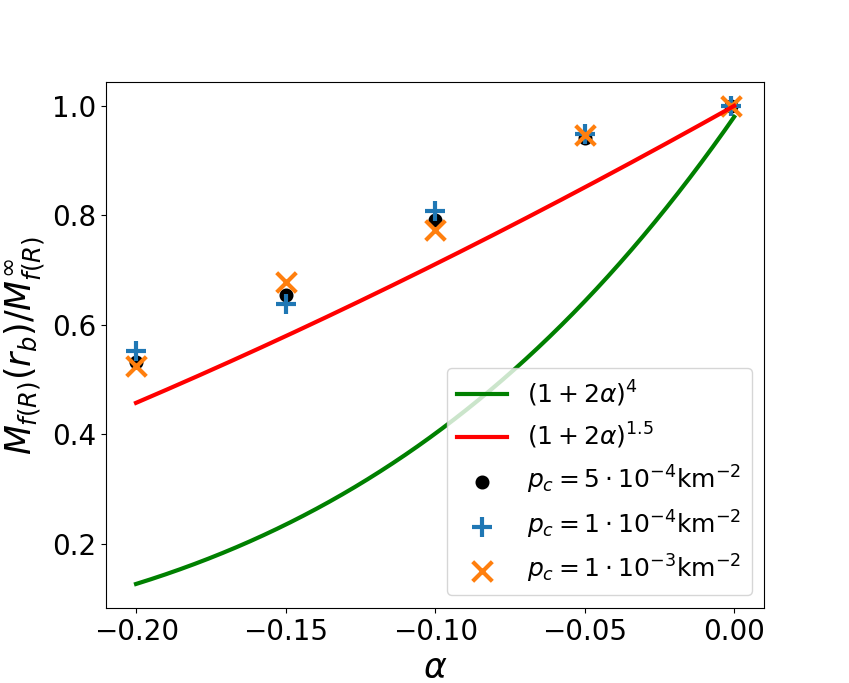}}
    \subfigure[]{
        \includegraphics[width=0.23\textwidth]{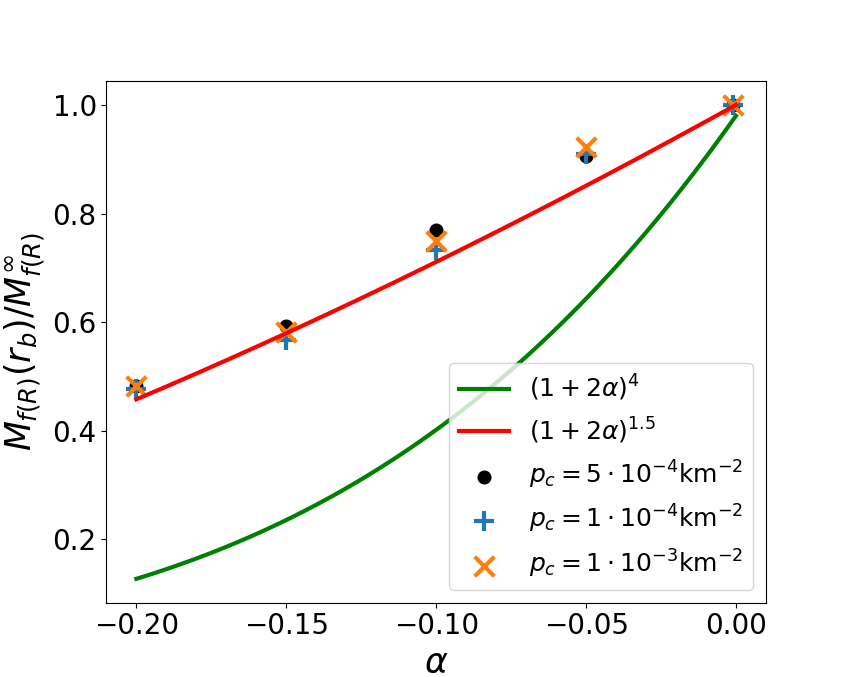}}
    \subfigure[]{
        \includegraphics[width=0.23\textwidth]{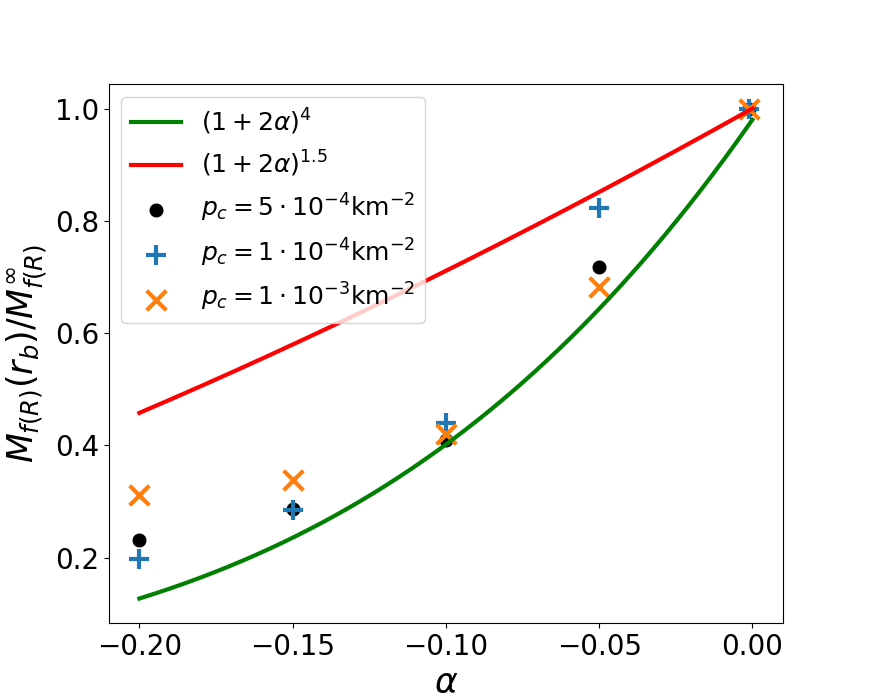}}    
    \caption{$M_{f(R)}(r_b)/M_{f(R)}^\infty$ results, as well as $(1+2\alpha)^4$ and $(1+2\alpha)^{1.5}$ curves as functions of $\alpha$. We consider three representative values for the central pressure, namely $p_c=5\cdot 10^{-4}, 1\cdot 10^{-4}, 1\cdot 10^{-3}\,{\rm km}^{-2}$ and the (a) stiff, (b) middle, (c) soft EOS under study.}
    \label{fig:Buch1_}
\end{figure}

To verify the validity of the functional in \eqref{eq:Buch_funcional}, for each of the three EOS under study, and for a range of central pressures, we compute $M_{f(R)}(r_b)/M_{f(R)}^\infty$ for various values of $\alpha$. The results are shown in Fig. \ref{fig:Buch1_} along with the curve $(1+2\alpha)^n$ for $n=1.5$ and $n=4$. Therein we observe that the curve for $n=1.5$ serves as a good bound for the middle and stiff EOS, but not for the soft one. However, the curve for $n=4$ is valid for all three EOS.

It is also noticeable that, for the soft EOS, the $M_{f(R)}(r_b)/M_{f(R)}^\infty$ quotient as a function of $\alpha$ is smaller than for middle, which in turn is smaller than for stiff. Thus, for a given value of $\alpha$, the slower the pressure grows with density, the smaller the $M_{f(R)}(r_b)/M_{f(R)}^\infty$ quotient yield. Consequently, since $(1+2\alpha)^4$ serves as a good bound for the soft EOS, we conclude that---for Starobinsky $f(R)$ quadratic models---the relation \eqref{eq:Buch_funcional} remains %in principle 
valid for any smooth EOS for which the pressure grows faster with density than in the soft case.

%\nocite{*}
\bibliography{bibliography}

\end{document}